\documentclass[useAMS,usenatbib,nofootinbib]{mn2e}
\usepackage{graphicx,color}
\usepackage[usenames,dvipsnames]{xcolor}
\usepackage{times}
\usepackage{natbib}
\usepackage{setspace}
\usepackage[pdftex]{hyperref}
\newif\ifAMStwofonts
\AMStwofontstrue

\newcommand{\be}{\begin{equation}}
\newcommand{\ee}{\end{equation}}
\newcommand{\ba}{\begin{eqnarray}}
\newcommand{\ea}{\end{eqnarray}}
\newcommand{\brr}{\begin{array}}
\newcommand{\err}{\end{array}}
\newcommand{\bc}{\begin{center}}
\newcommand{\ec}{\end{center}}

\def\Mpc{\, h^{-1} \, {\rm Mpc}}
\def\kMpc{\, h \, {\rm Mpc}^{-1}}
\def\Ms{\, h^{-1} \, M_\odot}

\newcommand{\msun}{M$_\odot$}
\newcommand{\pin}{{\sc pinocchio}}
\newcommand{\pinA}{{\sc P768}}
\newcommand{\pinB}{{\sc P3072-HR}}
\newcommand{\mice}{{\sc mice}}
\newcommand{\miceA}{{\sc MICE768}}
\newcommand{\miceB}{{\sc MICE3072-HR}}
\newcommand{\pthalos}{{\sc PTHalos}}

\newcommand{\mincir}{\raise
  -2.truept\hbox{\rlap{\hbox{$\sim$}}\raise5.truept \hbox{$<$}\ }}
\newcommand{\magcir}{\raise
  -2.truept\hbox{\rlap{\hbox{$\sim$}}\raise5.truept \hbox{$>$}\ }}
\newcommand{\siml}{\raise
  -2.truept\hbox{\rlap{\hbox{$\sim$}}\raise5.truept \hbox{$<$}\ }}
\newcommand{\simg}{\raise
  -2.truept\hbox{\rlap{\hbox{$\sim$}}\raise5.truept \hbox{$>$}\ }}

\title[Fast generation of DM halo catalogs] 
{An accurate tool for the fast generation of dark matter halo catalogs}

\author[P. Monaco et al.] 
 {P. Monaco,$^{1,2}$\thanks{E-mails: monaco@oats.inaf.it, emiliano.sefusatti@brera.inaf.it} E. Sefusatti,$^{3,4}$\footnotemark[1] 
 S. Borgani,$^{1,2,5}$ M. Crocce,$^{6}$
P. Fosalba$^{6}$,
 R. K. Sheth,$^{4,7}$ \and T. Theuns$^{8,9}$\\
$^1$ Dipartimento di Fisica - Sezione di Astronomia, Universit\`a di Trieste, via Tiepolo 11, I-34131 Trieste -- Italy\\ 
$^2$ INAF, Osservatorio Astronomico di Trieste, Via Tiepolo 11, I-34131 Trieste -- Italy \\
$^3$ INAF, Osservatorio Astronomico di Brera, Via Bianchi 46, I-23807 Merate (LC) -- Italy\\
$^4$ The Abdus Salam International Center for Theoretical Physics, Strada costiera, 11, I-34151 Trieste -- Italy\\
$^5$ INFN, Istituto Nazionale di Fisica Nucleare, Trieste -- Italy \\ 
$^6$ Institut de Ci\`encies de lÕEspai, IEEC-CSIC, Campus UAB, Facultat de Ci\`encies, Torre C5 par-2, Barcelona 08193 -- Spain\\
$^7$ Center for Particle Cosmology, University of Pennsylvania, 209 S. 33rd St., Philadelphia (PA) 19104 -- USA \\ 
$^8$ Institute for Computational Cosmology, Department of Physics, University of Durham, South Road, Durham DH1 3LE -- UK \\
$^9$ Department of Physics, University of Antwerp, Campus Groenenborger, Groenenborgerlaan 171, B-2020 Antwerp -- Belgium
}

\begin{document}

\maketitle

\label{firstpage}

\begin{abstract}
We present a new parallel implementation of the {\it PINpointing Orbit
  Crossing-Collapsed HIerarchical Objects} (\pin) algorithm, a quick
tool, based on Lagrangian Perturbation Theory, for the hierarchical
build-up of Dark Matter (DM) halos in cosmological volumes. To assess
its ability to predict halo correlations on large scales, we compare
its results with those of an N-body simulation of a 3 $h^{-1}$ Gpc box
sampled with $2048^3$ particles taken from the {\mice} suite, matching
the same seeds for the initial conditions. Thanks to the FFTW
libraries and to the relatively simple design, the code shows very
good scaling properties.  The CPU time required by {\pin} is a tiny
fraction ($\sim 1/2000$) of that required by the {\mice} simulation.
Varying some of {\pin} numerical parameters allows one to produce a
universal mass function that lies in the range allowed by published
fits, although it underestimates the {\mice} mass function of
Friends-of-Friends (FoF) halos in the high mass tail.  We compare the
matter-halo and the halo-halo power spectra with those of the {\mice}
simulation and find that these 2-point statistics are well recovered
on large scales. In particular, when catalogs are matched in number
density, agreement within ten per cent is achieved for the halo power
spectrum.  At scales $k>0.1\kMpc$, the inaccuracy of the Zel'dovich
approximation in locating halo positions causes an underestimate of
the power spectrum that can be modeled as a Gaussian factor with a
damping scale of $d=3\Mpc$ at $z=0$, decreasing at higher redshift.
Finally, a remarkable match is obtained for the reduced halo
bispectrum, showing a good description of nonlinear halo bias.  Our
results demonstrate the potential of {\pin} as an accurate and
flexible tool for generating large ensembles of mock galaxy surveys,
with interesting applications for the analysis of large galaxy
redshift surveys.
\end{abstract}


\section{Introduction}
\label{introduction}

Recent measurements of the Cosmic Microwave Background radiation
\citep[e.g.][]{BennettEtal2012, StoryEtal2012, DasEtal2013,
  Planck2013overview} have yielded accurate measurements of the
geometry of the Universe and the statistics of the linear, large-scale
perturbations visible at redshift $\sim1100$, the epoch of
recombination.  Thanks to these experiments, uncertainties on the main
cosmological parameters have been beaten down to the percent level
\citep{HinshawEtal2012, SieversEtal2013, Planck2013parameters}.  The
advantage of studying the Universe before perturbations started to
evolve beyond their linear regime is however counterbalanced by the
limit of observing it at a single cosmic time.  Performing
measurements at lower redshifts is then desirable because the
late-time growth of perturbations in a flat Universe is slowed down by
the dominance of the elusive dark energy, so measurements of the
growth of perturbations in the redshift range from $z=0$ to $z\sim1-2$
would lead to tight constraints on the equation of state of dark
energy and possibly provide evidence of physics beyond a simple
cosmological constant.  At the same time, accurate measurements of
density (through the galaxy power spectrum and higher moments),
potential (through galaxy weak lensing) and high density peaks
(through the mass function of galaxy clusters) can characterize the
growth of perturbations to a level of detail sufficient to distinguish
the predictions of General Relativity from those of some non-standard
gravity models \citep[e.g.][and references therein]{AmendolaEtal2012},
constrain other quantities like neutrino masses
\citep{LahavEtal2010,CarboneEtal2012,CostanziEtal2013} and the degree
of non-Gaussianity in the primordial perturbations
\citep{DesjacquesSeljak2010C, LiguoriEtal2010}.

For this reason many ongoing and future observational campaigns, such
as
DES\footnote{\href{http://www.darkenergysurvey.org/}{http://www.darkenergysurvey.org/}},
Euclid\footnote{\href{http://www.euclid-ec.org/}{http://www.euclid-ec.org/}},
PanSTARRS\footnote{\href{http://pan-starrs.ifa.hawaii.edu/public/science-goals/galaxies-cosmology.html}{http://pan-starrs.ifa.hawaii.edu/public/science-goals/galaxies-cosmology.html}},
LSST\footnote{\href{http://www.lsst.org/lsst/science}{http://www.lsst.org/lsst/science}}
or
SKA\footnote{\href{http://www.skatelescope.org/}{http://www.skatelescope.org/}},
are surveying or will survey large parts of the sky to a depth that
will reach $z\sim1$ or beyond.  Taking the future Euclid mission
\citep{LaureijsEtal2011} as an example, with $\sim15,000$ sq.  deg of
the sky surveyed in the $0.5<z<2$ redshift range, uncertainties in the
estimates of physical parameters from observable quantities will be
significantly affected by systematics connected to sample variance and
to the bias with which galaxies trace mass. This bias is ultimately
determined by the complex physics of baryons and will generally depend
on redshift and on the specific sample selection.  An accurate
assessment of these theoretical systematics requires the use of
numerical simulations to generate the non-linear distribution of Dark
Matter (hereafter DM) and models to populate the resulting DM halos
with mock galaxies.  Even assuming that large-scale structures can be
accurately described by the gravitational evolution of pure
collisionless DM and that the generation of galaxies in DM halos is
under control, the requirements for mock catalogs (typically of
Gpc$^3$ volumes and mass resolution below $10^{10}\Ms$ for on-going
and future experiments) are quite demanding.  Such large simulations
need more than $10^{10}$ particles, on-the-fly group finders and
nearly $100$ outputs to generate merger trees and past-light-cones. In
this case the hardware requirements in terms of memory and disc
storage raise more problems than the computing time needed to carry
out a single simulation.  The problem becomes untreatable when a very
large number of realizations (of order $1000$ or more) is needed to
estimate the covariance matrix of the galaxy power spectrum
\citep[e.g.][]{ManeraEtal2013}.  This is even more so for higher-order
statistics \citep{SefusattiEtal2006}.

This has prompted a number of recent works which use approximations to
the mildly non-linear evolution of perturbations
\citep[e.g.][]{ManeraEtal2013, KitauraHess2012,
  TassevZaldarriagaEisenstein2013}.  Several of these are based on
Lagrangian Perturbation Theory \citep[hereafter
  LPT,][]{MoutardeEtal1991,BuchertEhlers1993,Catelan1995}), a
perturbative solution of a set of equations for the displacements of
mass elements from their initial position.  With LPT it is possible to
accurately recover the large-scale density field of matter, but a fair
reconstruction of DM halos requires a different approach.

A decade ago, \citet[hereafter Paper I]{MonacoTheunsTaffoni2002}
presented a code, called {\it PINpointing Orbit Crossing-Collapsed
  HIerarchical Objects} (\pin), which was able to generate, with very
limited computing resources, a catalog of DM halos with known mass,
position and velocity from a realization of a Gaussian density
contrast field on a cubic grid, i.e. the same initial conditions that
are used by most simulations.  In that paper and in
\citet{TaffoniMonacoTheuns2002} the code was thoroughly tested against
two simulations that were state-of-the-art at that time.  It was shown
not only to reproduce (to within $\sim$5--20 per cent) statistics such
as the mass function and 2-point correlation function of DM halos, but
also to generate DM halos that agreed with the simulated ones at the
object-by-object level.  The code was tested by other groups, who
confirmed its accuracy in reproducing merger histories
\citep{LiEtal2007, ZhaoEtal2009} and velocities of DM halos
\citep{HeisenbergSchaferBartelman2011}.  It was also used by several
groups to study, e.g., DM halo density profiles \citep{LuEtal2006} and
concentrations \citep{ZhaoEtal2003}, the Sunyaev-Zeldovich effect in
clusters \citep{PeelBattyeKay2009} the properties of X-ray-selected
clusters \citep{PierreEtal2011}, galaxy clustering
\citep{ZhengCoilZehavi2007}, the formation of the first stars
\citep{SchneiderEtal2006} and of supermassive black holes
\citep{JahnkeMaccio2011}.

In this paper we present a new version of the {\pin} code, designed to
perform large runs (in our tests we use up to $2160^3$ particles) on
hundreds of computing cores of a parallel computer.  With respect to
\cite{MonacoTheunsTaffoni2002}, this version implements the same
algorithm but is fully parallel.  We test the accuracy of {\pin} on a
much larger range of masses and scales by comparing its results with a
large simulation kindly made available by the {\mice} collaboration
\citep{FosalbaEtal2008, CrocceEtal2010}.  We address clustering in
Fourier space, and demonstrate that the accuracy with which power
spectrum and bispectrum of DM halos is reconstructed can be easily
pushed below the $\sim10$ per cent level.  We also show CPU time
requirements and scaling properties to demonstrate that this code can
easily scale up to hundreds of cores, and identify the improvements
that are needed to run it on thousands of cores.  As an example, a
$2160^3$ realization requires 38 minutes of wall-clock time on 324 2.4
GHz cores of a linux machine, for a cost of 206 CPU hours, so that
running $10000$ such realizations would require just $2\times10^6$
CPU-hours.  This code provides fine time sampling of merger histories,
necessary to reconstruct halo positions along the past-light cone
\citep{MersonEtal2013, ManeraEtal2013} and to run semi-analytic models
of galaxy formation \citep{BensonEtal2012}.  This makes it invaluable
for addressing a range of problems such as sample variance, the
estimation of covariance matrices or sampling of parameter space,
where many very large realizations are needed.

The paper is organized as follows. Section 2 presents the algorithm,
its latest parallel implementation and its performance. Section 3
presents the simulations used for a comparison. In section 4 we
quantify the accuracy with which power spectrum and bispectrum of DM
halos are recovered.  Finally, Section 5 discusses the results and the
main conclusions.  The code is available under GNU/GPL license on
{\href{http://adlibitum.oats.inaf.it/monaco/Homepage/Pinocchio/index.html}
  {\small{
      http://adlibitum.oats.inaf.it/monaco/Homepage/Pinocchio/index.html}}}.

\section{PINOCCHIO}
\label{pinocchio}

The code is fully described in paper I, so here we only give a
brief account of how it works.

\subsection{The algorithm}
\label{algorithm}

The algorithm behind the {\pin} code has roots in the Extended Press
\& Schechter approach \citep{BondEtal1991} and in its extension to
non-spherical collapse by \citet{Monaco1995} and \citet{Monaco1997A},
but it does not use the Fokker-Planck approach based on sharp
$k$-space filtering. The calculation starts from the generation of a
linear density field on a regular grid, as done when generating the
initial conditions of an N-body simulation, and is divided in two
parts: (i) the computation of collapse times of individual particles,
performed by smoothing the density field on several scales and using
an ellipsoidal model based on LPT to compute individual times of
collapse; (ii) the fragmentation of the collapsed medium into distinct
objects, performed with an algorithm that mimics the hierarchical
build-up of DM halos.

\subsubsection{Collapse times}
\label{collapsetimes}

We start from a realization of a Gaussian field on a cubic
grid\footnote{The code can run on non-cubic grids as well.} of
$N^3$ vertices, assumed to have a physical size $L$.
This Gaussian field is assumed to represent a linear density contrast 
field, defined as the density contrast at a very early time $t_i$, 
linearly extrapolated to the present:

\be \delta_l({\bf q})= \frac{\delta({\bf q},t_i)}{D(t_i)}\, , \label{deltal}\ee

\noindent
where ${\bf q}$ is the {\it Lagrangian} coordinate of the mass
element, i.e. its initial position at $t=0$, and the growing mode
$D(t)$ is normalized to unity at $z=0$.  The power spectrum of
$\delta_l$ is given by the cosmological model and
$\langle\delta_l^2\rangle=\sigma_8^2$ when the field is top-hat
smoothed on a scale of 8 Mpc/h.  Following the EPS approach, the
density field is smoothed on a set of smoothing radii $R$.  This
is done with a Gaussian filter so the resulting trajectories are not
random walks but are highly correlated in $\sigma^2$.  
Smoothing radii are chosen so that the corresponding mass variances
are logarithmically spaced in intervals of 0.15 dex; typically from
10 to 20 smoothing radii are needed to sample the trajectories.

As described in \citet{Monaco1995}, at early times the evolution of a
mass element can be described as the evolution of an ellipsoid, whose
principal axes are given by the deformation tensor, i.e. the Hessian 
of the (peculiar) gravitational potential.  This is true at least 
until the ellipsoid collapses on its shortest axis.
The dependence of ellipsoid evolution on the background cosmology can 
be approximately factorized out by using the linear growing mode as
a time coordinate. In this case a very good approximation of this 
evolution can be obtained by using third-order LPT \citep{Monaco1997A}.
The argument can be reversed, so that ellipsoidal collapse can be 
considered as a truncation of LPT where all non-locality is given by 
the deformation tensor.  This allows one to treat the collapse of the 
first axis as an event of ``orbit crossing'', after which the LPT 
approach breaks down.  LPT is slow to converge in the case of a 
sphere, and this leads to an overestimate of collapse times for 
spherical peaks; to fix it \citet{Monaco1997A} found an empirical 
correction that reproduces the numerical solution of ellipsoidal 
collapse for quasi-spherical cases.  

For each smoothing radius the code performs a series of Fast Fourier
Transforms (FFTs) to compute the deformation tensor.  Then, for each
grid point, and using the ellipsoidal truncation of 3rd-order LPT and
its correction for quasi-spherical cases, the code computes the time 
$t_{\rm coll}({\bf q})$ at which the mass element at ${\bf q}$ is expected 
to collapse.  Using the growing mode as a time coordinate, the relevant 
quantity is the inverse of the collapse time of each mass element ${\bf q}$:
\be 
F({\bf q})=1/D(t_{\rm coll}({\bf q}))\, . \label{F} 
\ee
In the EPS language, for each grid point ${\bf q}$ we construct a
trajectory in the plane defined by mass variance of the smoothed field
$\sigma^2(R)$ and inverse collapse time $F({\bf q}; R)$.  If we used
spherical collapse we would have $F=\delta/\delta_c$, so the {\it
  absorbing barrier} at $\delta_c$, the linear density contrast at
which collapse is expected to take place, is replaced by a barrier
placed at the inverse of the time (the growing mode) at which DM halos
are requested.  When a mass element is predicted to collapse at the
smoothing radius $R$, it is interpreted as belonging to a halo of mass
at least $M=4\pi R^3/3\, \bar\rho$ ($\bar\rho$ being the average
matter density).  (The absorbing barrier construction prevents the
same mass element from being assigned to a halos with mass smaller
than $M$.)  In the same spirit, for each grid point ${\bf q}$ the code
records the highest value of $F$ along the trajectory, called $F_{\rm
  max}$, the associated smoothing scale $R_{\rm max}$, and the
velocity $v_{\rm max}$ at that position when smoothed on scale $R_{\rm
  max}$.  $F_{\rm max}$ is interpreted as the time at which, given the
mass resolution of the realization, the grid point is expected to
collapse, and $v_{\rm max}$ is computed from the first derivative of
the peculiar potential each time the $F_{\rm max}$ value is updated.

\subsubsection{Fragmentation}
\label{fragmentation}

The first part of the algorithm provides, for each grid point,
an inverse collapse time $F_{\rm max}$ and a velocity $v_{\rm max}$,
plus the smoothing radius $R_{\rm max}$ at which these have been
computed. With these it is possible to predict, at any time,
which regions of Lagrangian space have gone into orbit crossing
collapse. The fragmentation of the collapsed medium into distinct DM
halos is done with a code that mimics the hierarchical clustering of
DM halos \citep[see also the description
  in][]{HeisenbergSchaferBartelman2011}.  

It is convenient to describe grid points as ``particles'' in the 
following.  One thing worth stressing is that orbit crossing collapse 
does not imply that the particle belongs to a DM halo, because the 
filamentary network lying outside the halos may have undergone such 
a collapse and yet be far from a fully relaxed state; the code makes 
a distinction between collapsed particles in halos and those in the 
filamentary network.
Particles are first sorted in descending $F_{\rm max}$, and considered
in this (effectively chronological) order.  For each collapsing particle 
the six neighboring particles are considered, and the different halos to
which the neighboring particles belong are counted.  The following
cases are possible.
\begin{itemize}
\item[(i)] All six neighbors have not yet collapsed. The particle is 
  then a peak of the $F_{\rm max}$ field and is treated as a new DM halo 
  with one particle.
\item[(ii)] The particle touches only one halo.  To decide whether the
  particle is to be accreted on it, both the particle and the halo are
  displaced using the Zel'dovich approximation (ZA, 1st-order LPT). 
  If $d$ is their distance after displacement, accretion takes place if 
  the particle gets ``near enough'' to the group:
  \be d < f_a \times R + f_{ra} + f_s (R\sigma(R))^{1.7} , \label{accretion}\ee
  where $R=\sqrt{N_h} \times 3/4\pi$ is the halo Lagrangian radius in grid
  units ($N_h$ being the number of particles belonging to it), $f_a$,
  $f_{ra}$ and $f_s$ are free parameters and the factor
  $(R\sigma(R))^{1.7}$ (with $\sigma(R)$ computed at the collapse
  time), discussed in Appendix A of paper I, is a correction for the
  increasing inaccuracy of Zel'dovich displacements as the density
  fields becomes more non-linear.  If the particle does not accrete
  onto any halo, it is tagged as a ``filament'' particle.  
  After each accretion event all neighboring particles that have been
  previously tagged as filaments are accreted onto the halo as well.
\item[(iii)] The particle touches more than one halo.  First the code 
  checks whether the particle should be accreted onto one halo (the one 
  with minimal value of $d/R$).  Then it checks each pair of halos to 
  determine whether they should be merged together.  This happens if 
  they get ``near enough'' after Zel'dovich displacements have been 
  applied:
  \be d < f_m \times {\rm max}(R_1,R_2) + f_{ma}\, ; \label{merging} \ee
  In case the particle was not supposed to accrete onto any halo, 
  accretion is checked again after the merger(s).
\item[(iv)] The particle touches only filament particles.  
  Then it is tagged as filament as well.
\end{itemize}

This fragmentation code allows a very accurate time sampling of the 
merger trees, because halos are updated each time a collapsing particle 
touches them.  The full catalog of DM halos is output each time it is 
requested, with masses, centers of mass in the Lagrangian space, 
displacements obtained with the ZA and peculiar velocities.  
Merger histories are output only at the final time, giving a complete 
time sampling by reporting the masses of merging halos at each merger. 
The values of the free parameters are chosen by fitting to the desired 
halo mass function.  Table~\ref{tab:parameters} lists the values used 
in this paper, and the effect that each parameter has on the mass 
function.

\begin{table}
\centering
\begin{tabular}[t]{l|c|c|l}
\hline
\hline
parameter & eq. & value & effect on mass function \\
\hline\hline  

$f_a$    & \ref{accretion} & 0.285  & normalization and slope   \\
$f_{ra}$ & \ref{accretion} & 0.180  & normalization   \\
$f_s$    & \ref{accretion} & 0.060  & dependence on mass resolution and $z$ \\
$f_m$    & \ref{merging} & 0.350   & slope   \\
$f_{rm}$ & \ref{merging} & 0.700  & abundance of poorely sampled halos \\
\hline
\hline
\end{tabular}
\caption{\small Adopted values of the best-fit parameters. The right
  column gives the effect that a change in that parameter has on the
  mass function.}
\label{tab:parameters}
\end{table}

\subsection{The code}
\label{code}

The original scalar code (Version 1) was written in Fortran 77 and
designed to work on a simple PC.  It allowed to perform runs of
$256^3$ particles on a 450 MHz PentiumIII machine with 512 Mbyte of
RAM in nearly 6 hours, a remarkable achievement that allowed to obtain
reasonable statistics of merger histories with no access to a
supercomputer.  Because memory is the limiting factor in this case,
the code has an out-of-core design: it keeps in memory only one
component of the derivatives of the potential at a time, while the
other components are saved on the disc.  The most time- and memory-
consuming part is the computation of collapse times, fragmentation
takes less than 10 per cent of time.

In 2005, P. Monaco and T. Theuns wrote a parallel
(MPI\footnote{Message Passing Interface}) version of {\pin} 
(Version 2), that was publicized among interested researchers and 
used in several of the papers mentioned in the introduction.  
It is written in Fortran 90 and uses the FFTW package 
\citep{FrigoJohnson2012} to compute Fast Fourier Transforms.  
While parallelizing the computation of $F_{\rm  max}$ is straightforward 
(FFTW takes care of most communications), the fragmentation code was 
parallelized rather inefficiently, with one task performing the 
fragmentation and other tasks acting as storage; fragmentation is so 
quick that even this parallelization gives reasonable running times.  
Memory requirements were still minimized with an out-of-core strategy.  
This code is suitable to run on tens of cores, and requires fast access 
to the disc; when the number of cores increases, reading and writing on 
the disc becomes the limiting factor.

The version we use in this paper (Version 3) has been designed to run
on hundreds if not thousands of cores of a massively parallel super-computer.  The two
separate codes have been merged and no out-of-core strategy is adopted, 
so the amount of needed memory rises by a factor of three with respect 
to the previous version.  The computation of collapse times is performed 
as in version 2.  Fragmentation is performed by dividing the box into 
sub-volumes and distributing one sub-volume to each MPI task.  The 
tasks do not communicate during this process, and each sub-volume 
needs to extend the calculation to a boundary layer, where 
reconstruction is affected by boundaries.  From our tests and for a 
standard cosmology, the reconstruction of the largest objects is
convergent when the boundary layer is larger than about 30 Mpc.  This
strategy minimizes the number of communications among tasks, and the
boundary layer requires an overhead that is typically of order of some
tens per cent for large cosmological boxes.  For small boxes at very
high resolution this overhead would become dominant, in which case the
serial code of Version 1 (on a large shared-memory machine) or the
parallel code of Version 2 would be preferable.

Because FFTW distributes memory to tasks in planes, while the
fragmentation code works with sub-boxes, a communication round is
needed between the two codes to redistribute $F_{\rm max}$ and
velocities.  In the version we use here we have implemented a naive
distribution scheme where tasks always communicate in pairs.

To generate the initial linear density field in the Fourier space, we
have merged {\pin} with a part of the code taken from {\sc N-GenIC} by
V. Springel\footnote{\href{http://www.mpa-garching.mpg.de/gadget/}{http://www.mpa-garching.mpg.de/gadget/}}.
Besides a few technical improvements with respect to the original
{\pin} code, this has the advantage to be able to faithfully
reproduce a simulation run from initial conditions generated with 
{\sc N-GenIC}, or with the second-order LPT (hereafter 2LPT) version by
R. Scoccimarro\footnote{\href{http://cosmo.nyu.edu/roman/2LPT/}{http://cosmo.nyu.edu/roman/2LPT/}}
\citep{CroccePueblasScoccimarro2006}, just from the knowledge of the
assumed cosmology and the random number seed.

The code has also been extended to consider a wider range of
cosmologies including a generic, redshift-dependent equation of state
of the quintessence, but the computation of collapse times based on
ellipsoidal collapse still relies on the assumption that the
dependence on cosmology is factorized out of dynamical evolution when
the growing mode $D(t)$ is used as a clock, an approximation that
should be tested before using the code for more general cosmologies.
Displacements of groups from their final position are still computed
with the ZA.

\subsection{Performance and scaling}
\label{scaling}

To test its performance and its strong and weak scaling properties, we
ran the code on the PLX machine at the Centro Interuniversitario del
Nord Est per il CAlcolo (CINECA), a linux infiniband cluster with each
node equipped by 2 six-core 2.4 GHz processors and 48 Gb of RAM.

\begin{figure*}
\centering{
\includegraphics[width=.98\textwidth]{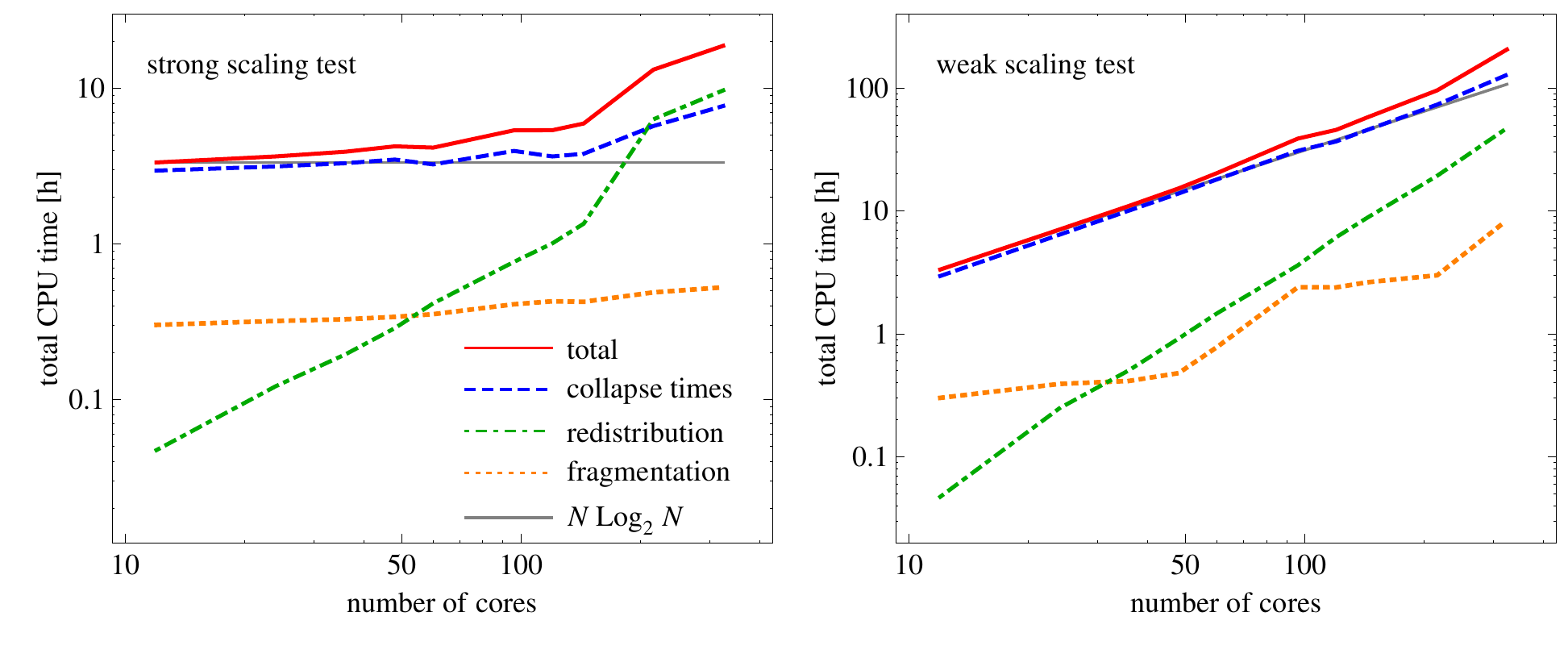}}
\caption{Left and right panels show a strong and a weak scaling test
  respectively.  In the strong test a 720 $\Mpc$ box of $720^3$
  particles is distributed on 1 to 27 nodes (12 to 324 cores, one task
  per core), while in the weak test the number of particles is
  increased proportionally to the number of cores, starting form the
  same simulation on a single node.
  In each panel we show the total CPU time needed to
  complete the run ({\em red, continuous curve}), to compute the
  $F_{\rm max}$ inverse collapse times ({\em blue, dashed, curve}), to
  redistribute the memory from planes to sub-volumes ({\em green,
    dot-dashed, curve}) and to fragment the collapsed medium into
  halos ({\em orange, dotted curve}). The black line gives the ideal
  $N\log_2N$ scaling (in the strong scaling test the number of
  particles $N$ is constant).}
\label{fig:scaling}
\end{figure*}

The left panel of Figure~\ref{fig:scaling} shows a strong scaling test
obtained by distributing a $720^3$ particles box of 720$\Mpc$ of
comoving length on 1 to 27 nodes, using 12 MPI tasks per node (one
task per core). We do not use multi-threading in this version of the code.
The red continuous curve gives the total time needed to complete the
run while the dashed blue, dot-dashed green and dotted orange curves
show, respectively, the time needed to compute inverse collapse times
$F_{\rm max}$, to perform the redistribution of memory from planes to
sub-boxes and to fragment the collapsed medium.  The horizontal black
line gives the ideal $N\log_2N$ scaling expected in this case 
(it is constant due to the fixed number of grid points).  
Thanks to the FFTW libraries,
the computation of collapse times scales very near the ideal case up
to 144 cores, with some increase of CPU time likely due to the
increased overhead of communications.  When more cores are used, FFTW
starts to distribute planes to tasks in an uneven way, so that only
some of the allocated cores are actually working (180 over 216 on 18
nodes, 240 over 324 on 27 nodes), while the others remain idle.  This
problem can be clearly avoided with a careful choice of the number of
tasks.  Fragmentation scales relatively well, with a modest increase
of CPU time related to the increasing overhead of boundary layers.
Redistribution is negligible for a small number of tasks but does not
scale; in this test it becomes dominant at the same time when collapse
times get far from the ideal scaling.

The right panel of the same figure shows a weak scaling test obtained
by increasing, at fixed mass resolution, the number of particles
proportionally to the number of cores used, up to $2160^3$ on 27
computing nodes \citep[the same number of particles as the Millennium
  simulation,][]{SpringelEtal2005}.  In this test we use rectangular
boxes.  The black line denotes the ideal $N \log_2 N$ scaling. Again,
computation of collapse times and fragmentation scale very near the
ideal case, while the redistribution becomes more and more significant
as the number of tasks increases, though it is not dominant even for
the largest simulation.

The $2160^3$ particles box run on 324 cores of the PLX machine takes
$\sim38$ minutes (for a cost of 206 CPU hours), with computation of
collapse times taking 62 per cent of time (37 per cent needed by
FFTs), redistribution 23 per cent and fragmentation 13 per cent.
Just to give an example, a numerical project of $10000$ 
Millennium-sized simulations on the same machine would require 
only $\sim2\times10^6$ CPU hours and would be over in less than 9 months
on only 324 core, or a month on about $3000$ cores.  An improvement of
the redistribution code would lower requirements by 20 per cent and
would allow the code to be applied to larger box sizes.

\section{Simulations}
\label{simulations}

To test the accuracy of {\pin} for the clustering of DM halos we 
compare to a simulation taken from the {\mice} suite of cosmological 
N-body simulations \citep{CrocceEtal2010}\footnote{Selected halo 
catalogs and other data are available for download at
  \href{http://maia.ice.cat/mice/}{http://maia.ice.cat/mice/}}. 
{\mice} is a large set of $\Lambda$CDM simulations performed with the 
{\sc Gadget}-2 code described in \citet{Springel2005}. The {\mice}
catalogs provide Friend-of-Friends (FoF, hereafter) halos with linking
length $b=0.2$ in units of the mean inter-particle distance. The
assumed cosmology is that of a flat, $\Lambda$CDM Universe with
$\Omega_m=0.25$, $\Omega_b=0.044$, $n_s=0.95$, $\sigma_8=0.8$ and
$h=0.7$ ($\Omega_b$ is used to generate the initial conditions but
all particles are collisionless).
In the rest of this paper we will use the term ``halos'' for 
both N-body FoF halos {\em and} for {\pin} ``groups'' of particles 
since this choice should not lead to any ambiguity.

We focus specifically on one run, {\miceB} \citep[following the
denomination of][]{CrocceEtal2010}, consisting of a box of sides 
$3072\Mpc$ sampled by $2048^3$ particles, each of mass $2.3\times10^{11}\Ms$.  
Note that, unlike other simulations in the {\mice} suite, the {\miceB} 
run does not use 2LPT initial conditions. We use here for the mass 
function a tabulated correction provided by the {\mice} collaboration 
for those runs with ZA initial conditions, but we do not attempt to 
correct halo masses when applying mass cuts to compute correlation 
statistics. The {\pin} run required 31 min on 25 computing nodes 
(300 cores), so the total cost was 155 CPU hours, a tiny fraction 
($\sim 1/2000$)
of
the $370,000$ hours needed by the N-body simulation on the Marenostrum 
supercomputer.
For testing purposes, we have also used the smaller 
{\miceA} simulation, a $768\Mpc$ box sampled by $1024^3$ particles with 
mass $2.9\times10^{10}\Ms$. This has a higher mass resolution, but its
volume is not large enough to be used for large-scale clustering statistics. 
For clarity, we use similar names for the {\pin} run, 
replacing the ``{\sc MICE}'' prefix with ``{\sc P}''.

The comparison with {\mice} simulations allows us to test {\pin} on much
larger volumes than previously done. The halo catalogs are public and 
the mass function analysis has been performed by the {\mice} collaboration
\citep{CrocceEtal2010}. As mentioned above, the current version of {\pin} 
makes use of the {\sc N-GenIC} code for the initial conditions, so it is 
simple to set up the same initial conditions used in the simulations, 
removing differences due to sample variance, and allowing a comparison 
at the object-by-object level. Extensions of {\sc NGenIC} to 2LPT initial 
conditions are available and are crucial for accurately simulating the 
halo mass function \citep{CroccePueblasScoccimarro2006}. Finally, the 
{\sc N-GenIC} code has been further extended to include non-Gaussian
initial conditions \citep{ScoccimarroEtal2012}. Although we will not
consider this possibility here, the extension of {\pin} to this
specific departure from the Standard Cosmological Model is, in
principle, straightforward.

\begin{figure*}
\begin{center}
\includegraphics[width=.98\textwidth]{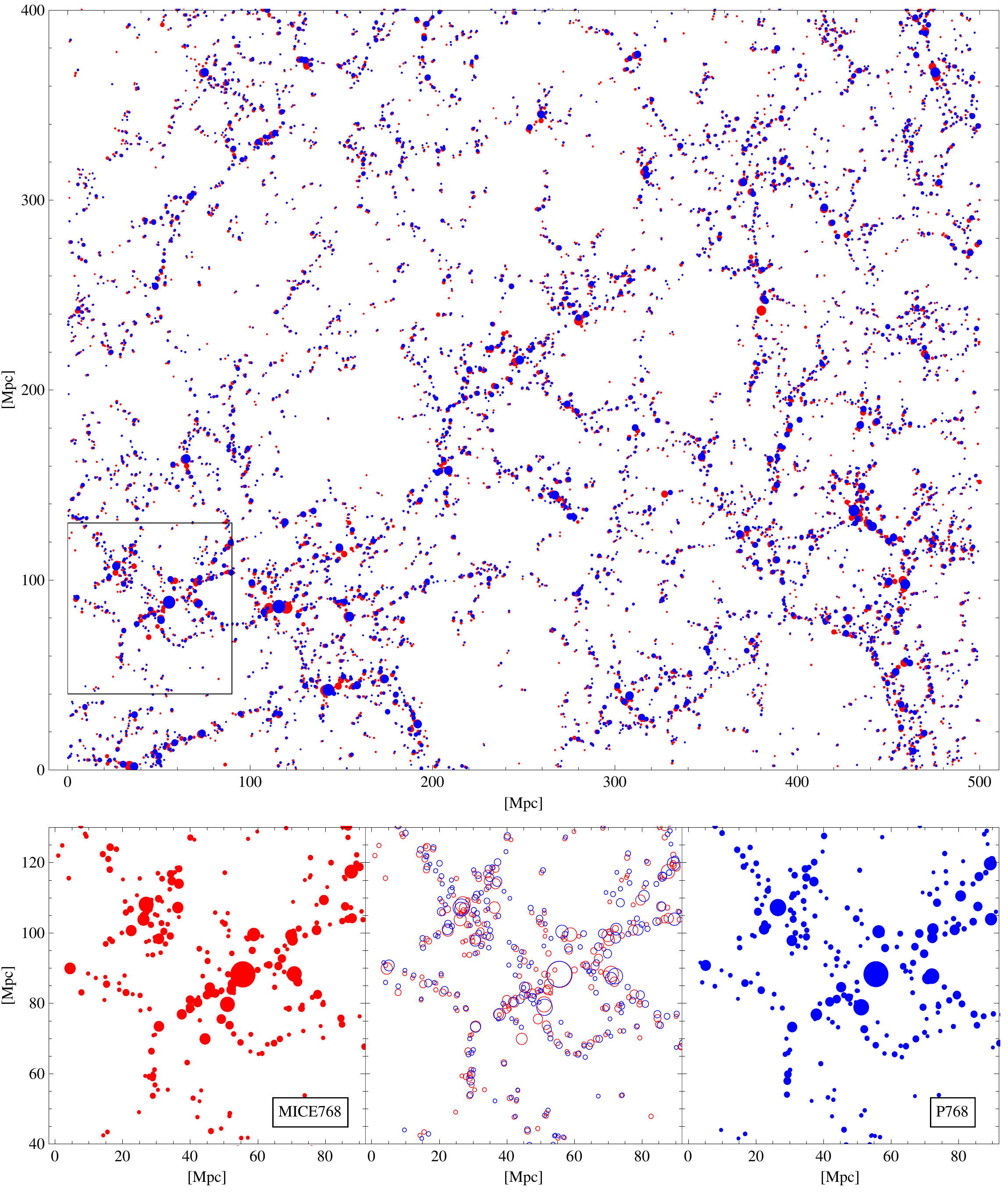}
\caption{\label{fig:slice500} Comparison between the halo
  distributions predicted by the {\pin} {\pinA} realization ({\em
    blue}) and the {\miceA} N-body simulation ({\em red}) on a
  $500\Mpc\times 400\Mpc$ field, $20\Mpc$ slice. The upper panel shows
  the entire field. The lower panels show a $70\Mpc\times 70\Mpc$
  zoom with the two separate distributions and overlapping as
  circles. All halos with $\log_{10} M/(\Ms)\ge 12.5$ are shown 
  by disks and circles having $1.7\times$ the virial radius. }
\end{center}
\end{figure*}

Fig.~\ref{fig:slice500} provides a first, qualitative comparison of
{\pin} with an N-body run at $z=0$.  For this comparison we have used 
the smaller {\miceA} simulation, that has a better mass resolution. 
In the large, top panel, corresponding to a $500\times400\times 20\Mpc^3$ 
volume, blue dots represent individual halos from {\pinA} plotted on 
top of the corresponding halos from {\miceA}, shown as red dots. 
The size of each dot is proportional to the halo virial radius. 
It has been enlarged for clarity, leading to unrealistic overlaps 
between halos in each realization. All halos with
$\log_{10} M/(\Ms)\ge 12.5$ are shown. The lower panels show in detail
a sub-volume of $90\times 90\Mpc$ area and same thickness of $20\Mpc$,
with the left and right ones corresponding respectively to the
individual N-body and {\pin} outputs and with the central one showing
again the two together by means of open circles to provide a clearer
comparison of sizes and positions.

While large-scale structure is well reproduced, it can be seen that 
the most massive halos in {\pin} tend to be more isolated than their  
{\mice} counterparts, which have more numerous smaller halos in their 
vicinity. We know from paper I that {\pin} provides a good a match  
at the object-by-object level, so this mismatch is related to the 
limitations of the ZA to properly reproduce the displacement field and 
reconstruct large-scale high density peaks.  The relatively thin slicing 
causes some matching halo pairs to be in or out the slice, thus 
artificially increasing the number of apparently unmatched halos.  
To make full sense of this comparison, one should take into account 
that a ``{\pin} halo'' does not exactly coincide with an FoF halo, 
though parameters can be tuned to maximize their similarity.  
These issues will be explored in detail elsewhere.

\subsection{Mass function and parameters}

As explained in section~\ref{fragmentation}, the construction of halos
in the {\pin} code depends on five free parameters whose values were 
determined in Paper I by fitting the mass function to the one obtained 
from simulations available at that time\footnote{A standard CDM simulation 
with $360^3$ particles in a box $500\Mpc$ on a side, and a smaller 
$\Lambda$CDM simulation with $256^3$ particles in a $100\Mpc$ box.}.  
The mass function was shown to be accurate at the 5 per cent level when 
compared to FoF halos with linking length $b=0.2$ times the inter-particle 
distance, though a 10-20 per cent underestimate at large masses and 
high redshifts was reported.

For a proper comparison with much bigger simulations the parameters
can be retuned to improve the agreement to a higher level of accuracy.  
To fully and properly complete such a task we must perform a large 
number of realizations and a detailed study at the object-by-object 
level of {\pin} halos in comparison with FoF and Spherical Overdensity 
(SO, hereafter) halos, paying specific attention to the high-mass tail. 
This will be presented in a forthcoming paper. The present paper aims 
at presenting a first test of version 3 of the {\pin} code on cosmological
volumes characterized by box sizes and particle numbers that are more 
than two orders of magnitude larger than those previously addressed.

\begin{figure}
\begin{center}
\includegraphics[width=.48\textwidth]{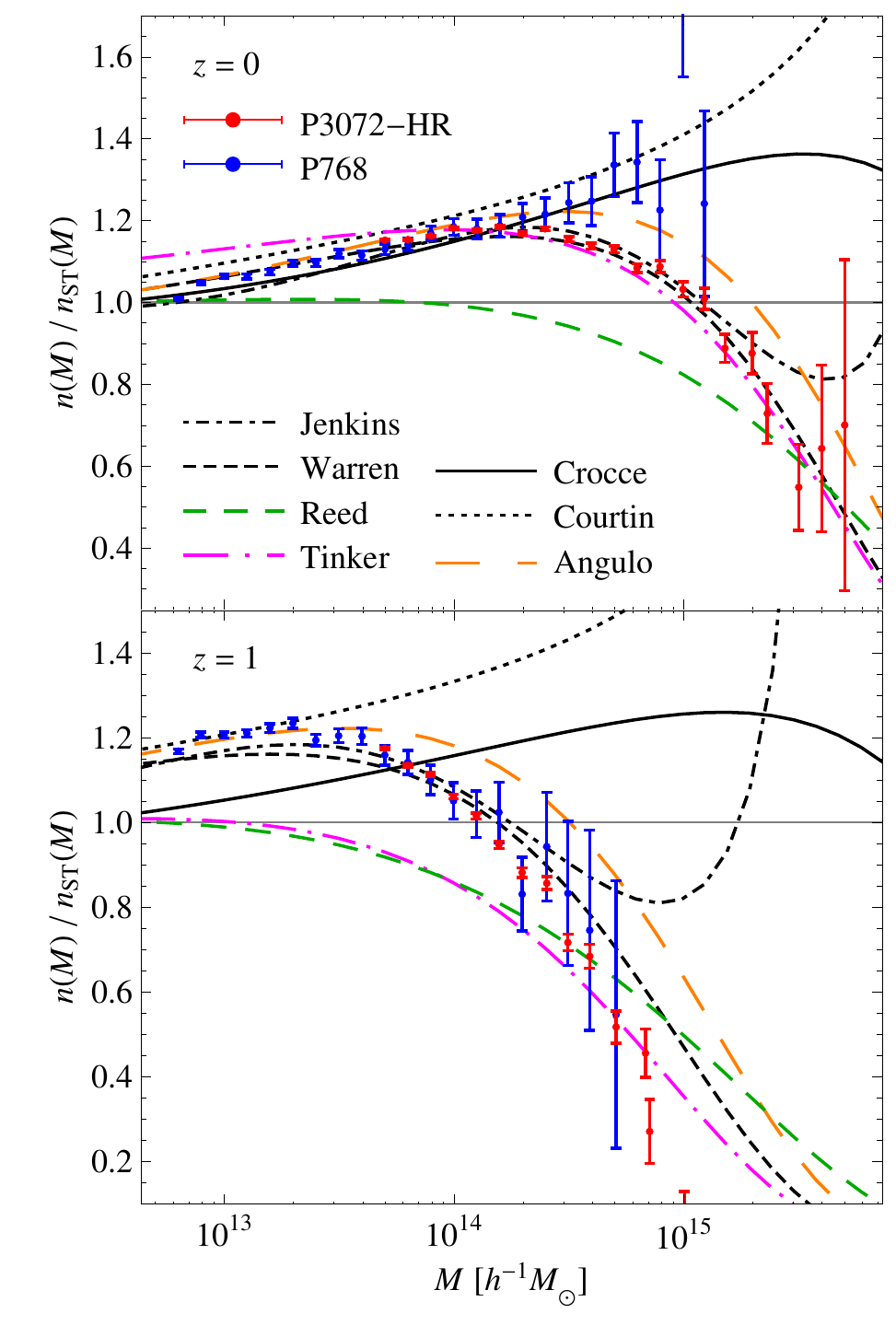}
\caption{\label{fig:fits} Ratio of the mass function measured in the {\pinA} ({\em blue data points}) and {\pinB} ({\em red data points}) realizations to the \citet{ShethTormen1999} fitting formula. For comparison, the same ratio is shown for several other analytic fits taken from the literature: \citet{JenkinsEtal2001} ({\em black, dot-dashed curve}), \citet{WarrenEtal2006} ({\em black, short dashed}), \citet{ReedEtal2007} ({\em green, medium dashed}), \citet{TinkerEtal2008} ({\em magenta, dot-long dashed}), \citet{CrocceEtal2010} ({\em black, continuous}), \citet{CourtinEtal2011} ({\em black, dotted}) and \citet{AnguloEtal2012} ({\em orange, long dashed}).}
\end{center}
\end{figure}

At $z=0$, numerical convergence among mass functions of simulated DM 
halos has been reached at the $\la5$ per cent level for masses 
$\la 10^{14}$ {\msun}.  For larger masses, differences among simulations 
can amount to several 10s of per cent.  At fixed mass, the disagreement 
worsens at higher redshift, where objects correspond to rarer peaks of 
the linear density field.  This is also an effect of the steepness of 
the high-mass tail of the mass function, because of which small 
differences in mass result in large differences in number density.  
Moreover, the mass function is approximately ``universal'', i.e. mass 
functions at all redshifts lie on the same relation when the adimensional 
quantity $(M^2/\bar\rho)(d n(M)/d M)$ 
($dn(M)$ being the number density of halos of mass between $M$ and $M+dM$) 
is shown as a function of $\nu=\delta_c/\sigma(M)$, with $\sigma(M)$ 
being the mass variance at the scale $M$. However, recent determinations 
have reported small but significant violations of universality 
\citep{CrocceEtal2010,TinkerEtal2008}.
Figure~\ref{fig:fits} compares, in terms of ratios to the 
\citet{ShethTormen1999} fitting formula, 
several analytic fits from the literature, obtained 
both for FoF \citep{JenkinsEtal2001, WarrenEtal2006, ReedEtal2007, CrocceEtal2010, CourtinEtal2011, AnguloEtal2012} and SO \citep{TinkerEtal2008} halos. 

Using {\pin} with the parameter values given in paper I, we
confirmed the tendency, already noticed in paper I and in
\cite{PeelEtal2009}, of {\pin} to underestimate the number density of
rare objects.
To improve this trend we performed some parameter tuning.  We
  first dropped the dependence, described in Appendix A of paper I, of
  $f_a$ and $f_{ra}$ on resolution. Both $f_a$ and $f_{ra}$ influence
  the normalization of the mass function, but the first one also
  steepens it. We checked that increasing $f_a$ to 0.285 while
  lowering $f_{ra}$ to 0.180 provides a number density of rare
  objects compatible with simulations though at the lower end of the
  allowed range.  The
  other parameters were left as in paper I; Table~\ref{tab:parameters}
  reports the parameters values used in this paper.  Finer tuning can
  be achieved by using several large nested boxes and sampling a wider
  parameter space; because the number of constraints is much larger
  than the number of parameters, degeneracies in parameter values can
  be broken with this approach.
In Figure~\ref{fig:fits} we report
the mass function measured 
in the {\pinA} ({\em blue data points}) and {\pinB} ({\em red data points}) 
realizations, 
the error bars on the data points represent simply the expected Poisson error.
We notice in the first place good agreement between the results for
the two boxes with different resolution, particularly at large
redshift. 
We find good agreement over quite a
large range of masses with the \cite{WarrenEtal2006} fit \citep[and to
  a lesser extent to the][one]{AnguloEtal2012} 
However, we were unable to reproduce the
{\mice} FoF counts of \cite{CrocceEtal2010} at high redshift. Since
the agreement with the SO mass function of \cite{TinkerEtal2008} is
better, the disagreement with {\mice} may related to the tendency of
FOF to overlink halos and to include filaments that surround the
rarest halos.

\begin{figure}
\begin{center}
\includegraphics[width=.48\textwidth]{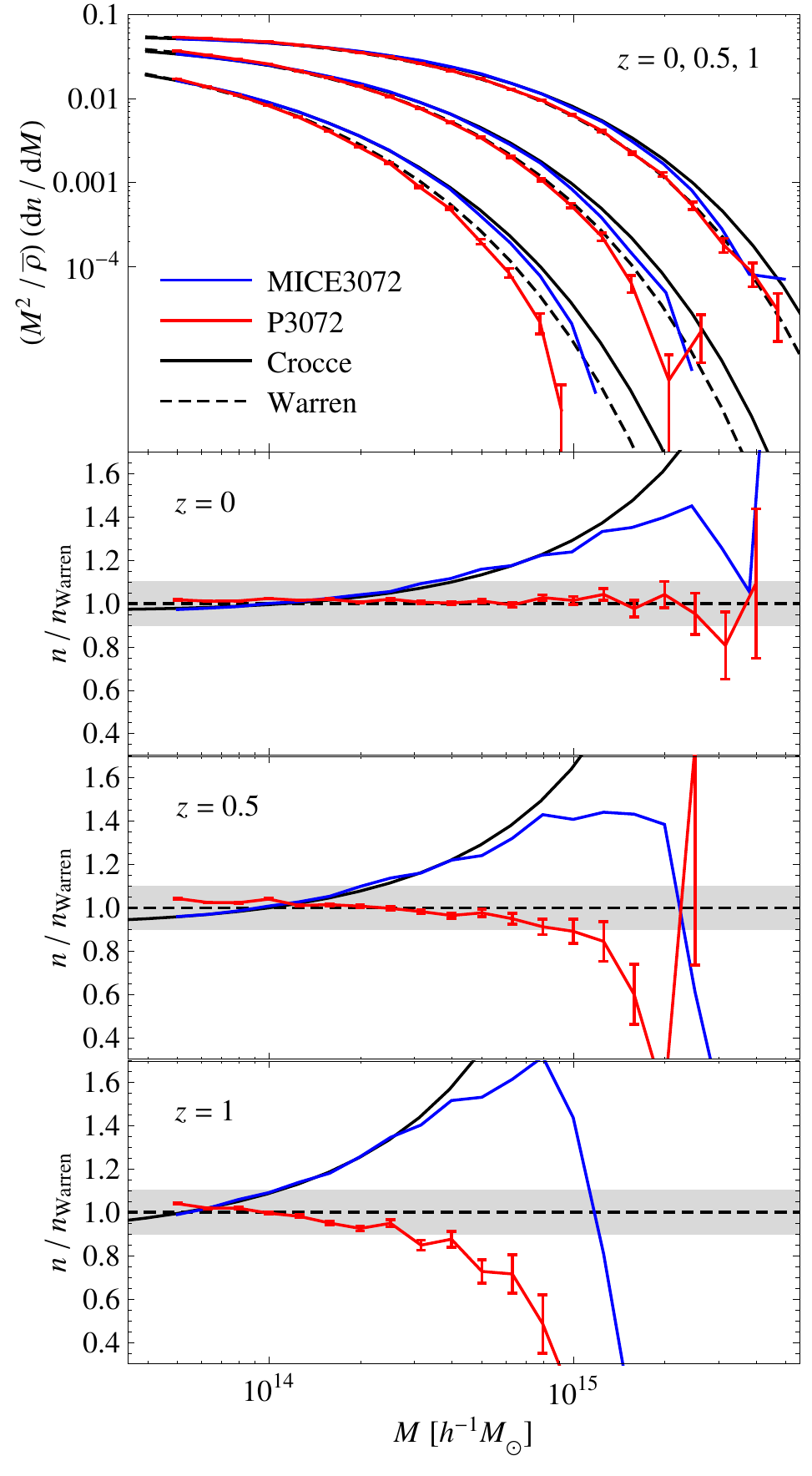}
\caption{\label{fig:mf} The top panel shows the adimensional mass function predicted by {\pin} ({\em red, continuous curve with error bars}) compared with the \citet{WarrenEtal2006} fit ({\em black, dashed curve}) and the {\mice} fit ({\em black, continuous curve}) and data ({\em blue, continuous curve}) for {\miceB} at $z=0$, $0.5$ and $1$. The lower panels show, for each redshift, the residuals w.r.t. the \citet{WarrenEtal2006} fit with, in addition to the {\mice} results.
The shaded gray region corresponds to deviations within 10\% w.r.t. the \citet{WarrenEtal2006} prediction.}
\end{center}
\end{figure}

To assist in the interpretation of the correlations results presented in the next section, Fig.~\ref{fig:mf} shows a more direct comparison of both the {\mice} fit and the {\miceB} mass function with the one from {\pinB}. In particular, the upper panel of Fig.~\ref{fig:mf} shows the {\pin} adimensional mass function ({\em red curve with error bars}) with the results of {\miceB} ({\em blue curve}) and the analytic fits of \citet{CrocceEtal2010} ({\em black curve}) and \citet{WarrenEtal2006} ({\em black, dashed curve}). The lower panels give the residuals for both {\pin} and {\mice} results w.r.t. the \cite{WarrenEtal2006} fitting formula. Error bars are shown only for the {\pin} output and account for Poisson noise. 

The lower mass bin used in this plot corresponds to halos of a minimum of 200 particles, both for {\pin} and {\mice}, i.e. the same cut-off assumed in \citet{CrocceEtal2010} for the fitting procedure. FoF masses from the {\mice} catalogs account for the mass correction suggested by \citet{WarrenEtal2006} in order to avoid the statistical noise effects due to the estimate of the halo density field with a finite number of particles.  This correction corresponds to defining a ``corrected'' number of particles per halo given by $ n_p^{corr}=n_p\,(1-n_p^{-0.6})$.  Also, the {\mice} mass function is corrected as suggested in \cite{CrocceEtal2010} to reproduce the result of 2LPT initial conditions.  We do not consider such corrections for {\pin} masses, which are instead given simply by the number of particles belonging to a given halo.

As already noted, {\pin} reproduces the mass function of {\miceB} within 10 per cent at $M\sim 10^{14}\Ms$.  At larger masses, however, it increasingly underestimates the {\mice} halo number density, especially when compared with the {\mice} analytic fit that takes advantage of the results from the full {\mice} set, including two boxes of larger size with respect to {\miceB}.  Thanks to the parameter tuning, the {\pinB} mass function reproduces the Warren fit to within a few per cent in the range where the {\miceB} mass function closely follows the {\mice} analytic fit.  For $z>0$ the {\pin} mass function goes below the \cite{WarrenEtal2006} fit, but this happens in the same mass range where the {\mice} mass function for the same {\miceB} box starts to underestimate the analytical fit obtained using larger boxes. So this discrepancy may be related to the smallness of the box.

\section{Accuracy tests for clustering statistics}
\label{accuracy}

In this section we present a direct comparison of the halo power
spectrum and bispectrum predicted by the current version of {\pin}
with their counterparts measured in the {\miceB} simulations.

\subsection{Power spectrum}
\label{powerspectrum}

Paper I, using the two-point correlation function, showed that on the
relatively small scales ($<30\Mpc$) testable with those simulations
and using Zel'dovich displacements, the clustering of halos is
recovered by {\pin} at the 10-20 per cent level.  Here we examine
instead measurements of the halo-halo power spectrum and the 
halo-mass cross-power spectrum on significantly larger scales, 
encompassing the acoustic oscillation range at low cosmic variance.

\begin{figure}
\begin{center}
\includegraphics[width=.48\textwidth]{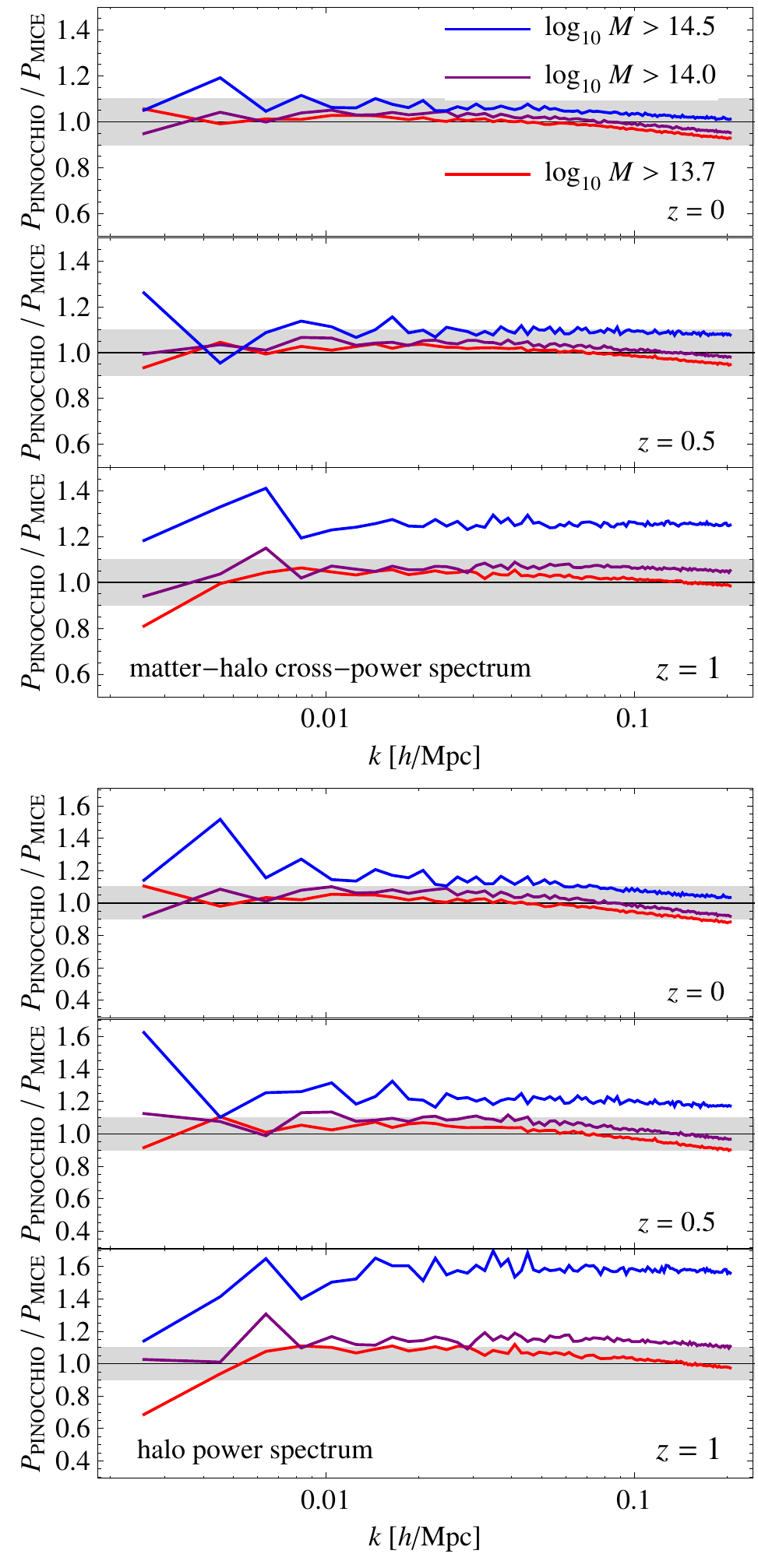}
\caption{\label{fig:psratio} Ratios between power spectra computed using the {\pinB} and {\miceB} catalogs. Top panels show the matter-halo cross-power spectrum, bottom panels the halo-halo power spectrum. We show results at redshifts $z=0$, $0.5$ and $1$ for different thresholds in mass defined by $\log_{10} (M\Ms) > 13.7$, $14$ and $14.5$. The shaded gray area corresponds to discrepancies below 10\%.}
\end{center}
\end{figure}

A first comparison between {\pin} and {\mice} is performed in mass thresholds, taking each halo mass at face value, as predicted by {\pin}, without any rescaling to match the two mass functions. Therefore, any mass function discrepancy will affect the normalization of the power spectrum.  Fig.~\ref{fig:psratio} shows the ratios between the matter-halo ({\em top panels}) and halo-halo ({\em bottom panels}) power spectra from {\pinB} and the corresponding ones from the {\miceB}. The first quantity, in particular, is obtained by correlating the halo number density, computed on a grid with a cloud-in-cell algorithm, with the non-linear mass density field measured from the simulation output for both {\pin} and {\mice}.  We consider as an example three distinct populations defined by $\log_{10} (M/\Ms) > 13.7$, $14$ and $14.5$, with $M$ measured in. The lowest mass threshold corresponds to halos of 200 particles.  Finally, for the halo power spectrum comparison we keep the shot-noise contribution, since the halo power spectrum including shot-noise is more relevant for covariance estimation purposes.

Differences in the mass functions will result in differences in the power spectrum (mostly its normalization), that will be larger for the halo-halo case. For a fixed mass threshold, in fact, {\pin} objects are relatively rarer, and therefore more biased, than their {\mice} counterparts. The results of Fig.~\ref{fig:psratio} show that for the lowest mass thresholds ($\log_{10}(M/Ms)> 13.7$) and lower redshift $z\le 0.5$, where the mass function is well-matched, the discrepancy between the {\pin} and {\mice} cross-power is below the 5\% level at scales $k<0.1\kMpc$. The agreement worsens in the case of the halo-halo power spectrum but stays well within the 10\% level. Larger discrepancies in the normalization, of order of 10-20 per cent, occur for the largest mass threshold and at $z=1$. 

\begin{figure}
\begin{center}
\includegraphics[width=.48\textwidth]{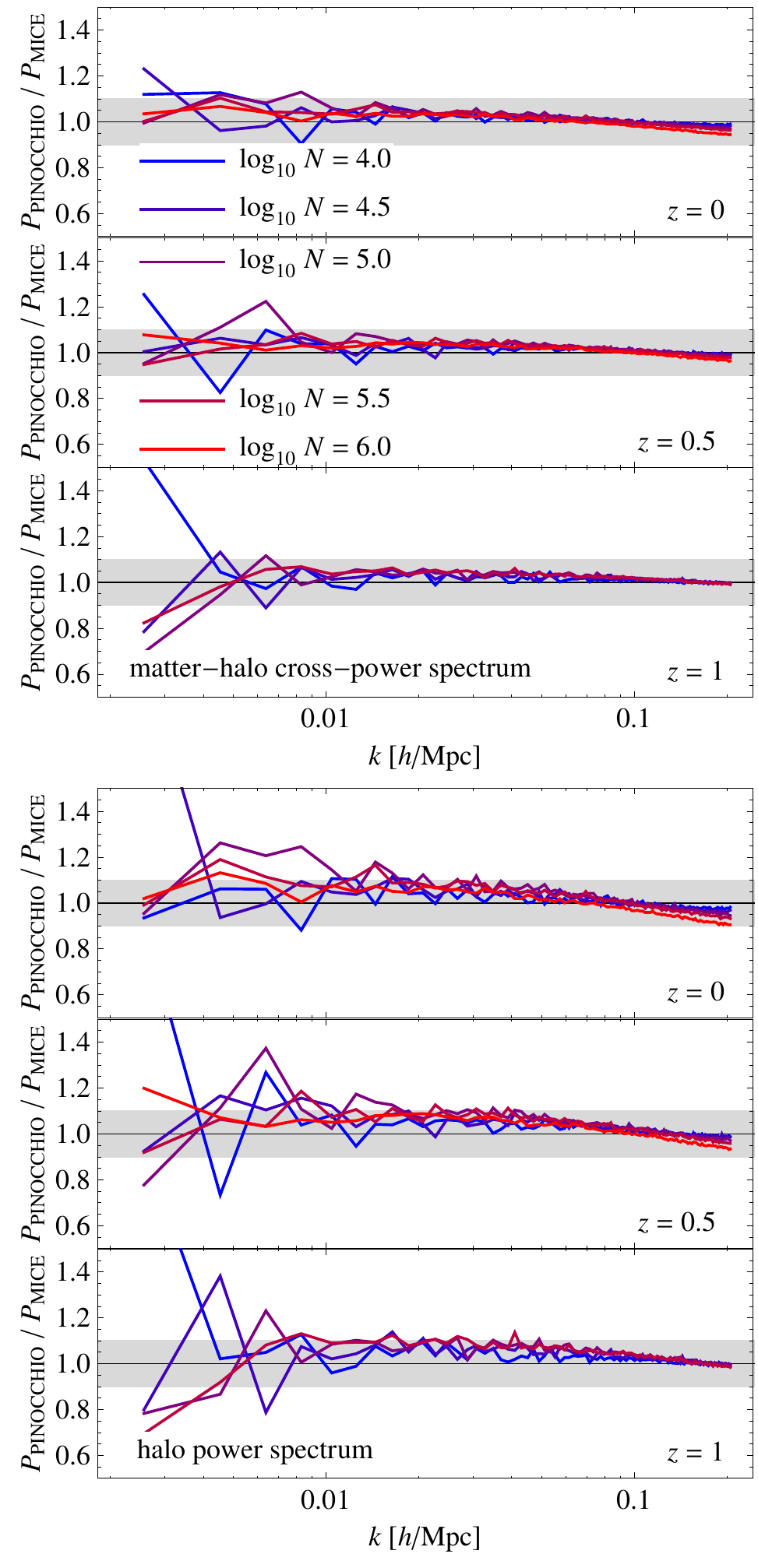}
\caption{\label{fig:psratiodens} Same as Fig.~\ref{fig:psratiodens} but for halo populations defined by a fixed number $N$ of the most massive halos with $\log_{10}N=4$, $4.5$, $5$, $5.5$ and $6$. The shaded gray area correspond to discrepancies below 10\%.}
\end{center}
\end{figure}
To confirm our interpretation of the impact of the mass function discrepancy on
the power spectrum, we consider as well power spectrum measurements performed on halo populations defined directly in terms of halo number density. The corresponding ratios with the {\mice} results are shown in Fig.~\ref{fig:psratiodens}. In this case we consider populations defined by a total number of most massive halos $N$ taking the values $\log_{10}N=4$, $4.5$, $5$, $5.5$ and $6$, corresponding to masses roughly ranging from $\log_{10}(M/\Ms)\simeq 13.9$ to $14.9$ at redshift zero.  Notice the remarkably low scatter among different density populations. The overall departure from the {\mice} results at large scales is about $4$ and $8\%$ for the cross- and halo-halo power spectra, respectively. 

\begin{figure}
\begin{center}
\includegraphics[width=.48\textwidth]{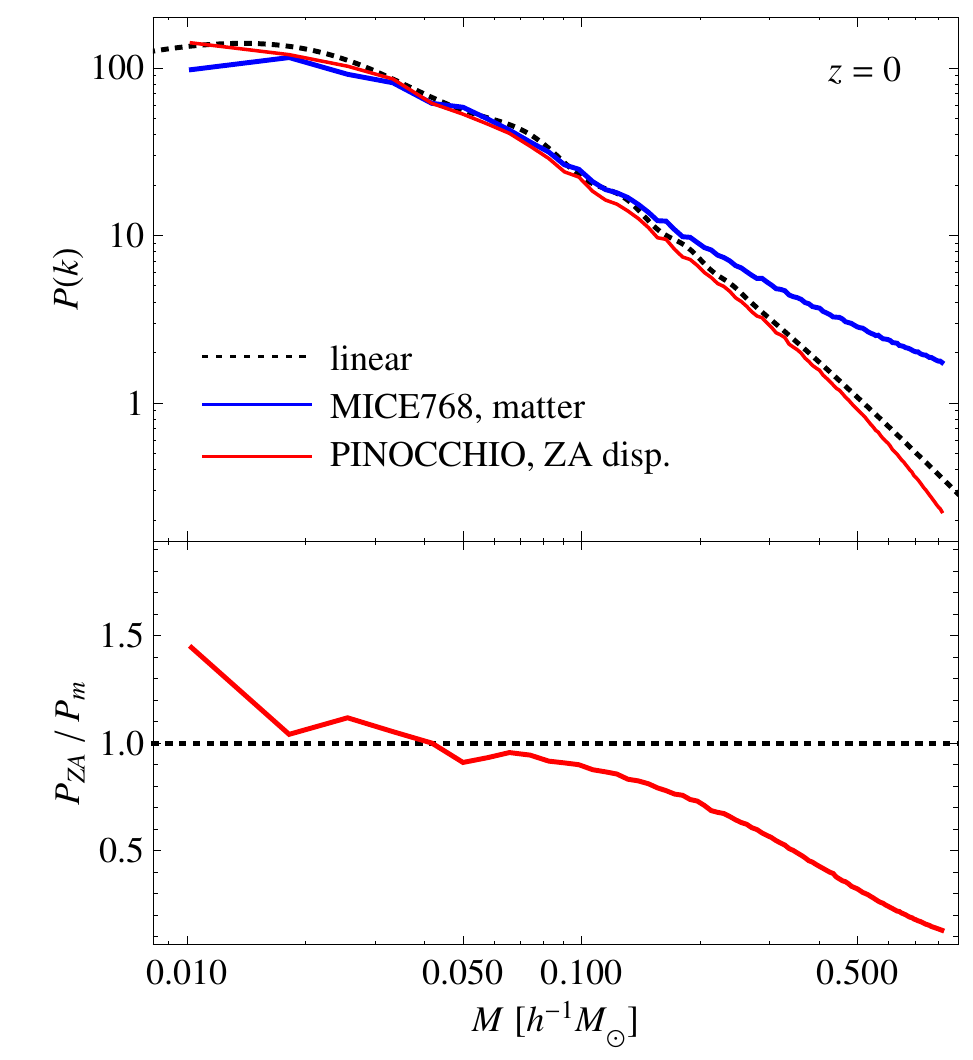}
\caption{\label{fig:ZApowmatter} 
{\em Top panel}: Comparison of the nonlinear matter power spectrum from
{\miceA} ({\em blue curve}) with that measured using ZA displacements
for all particles ({\em red, continuous curve}), and in linear theory 
({\em dotted curve}). {\em Bottom panel}: ratio between the {\pin} ZA 
 power spectrum and the fully nonlinear matter power spectrum in {\mice}.}
\end{center}
\end{figure}

In the mildly nonlinear regime an increasing suppression of {\pin} power spectra is also evident.  It is stronger for smaller halos, and is likely related to {\pin}'s use of the ZA for particle displacements and consequent halo positions.  To demonstrate this, Fig.~\ref{fig:ZApowmatter} shows the ratio between the matter power spectrum obtained from the Zel'dovich displacements for all particles and the same quantity computed using the simulation output. We consider, in this case, the smaller {\miceA} run and the corresponding {\pinB} one. Similarly to Fig.~\ref{fig:psratio}, this ratio shows a damping of the ZA power spectrum by 15 per cent at $k=0.1\kMpc$.  Clearly {\pin} halos cannot show better clustering properties as long as ZA is used for the displacements.

\begin{figure}
\begin{center}
\includegraphics[width=.48\textwidth]{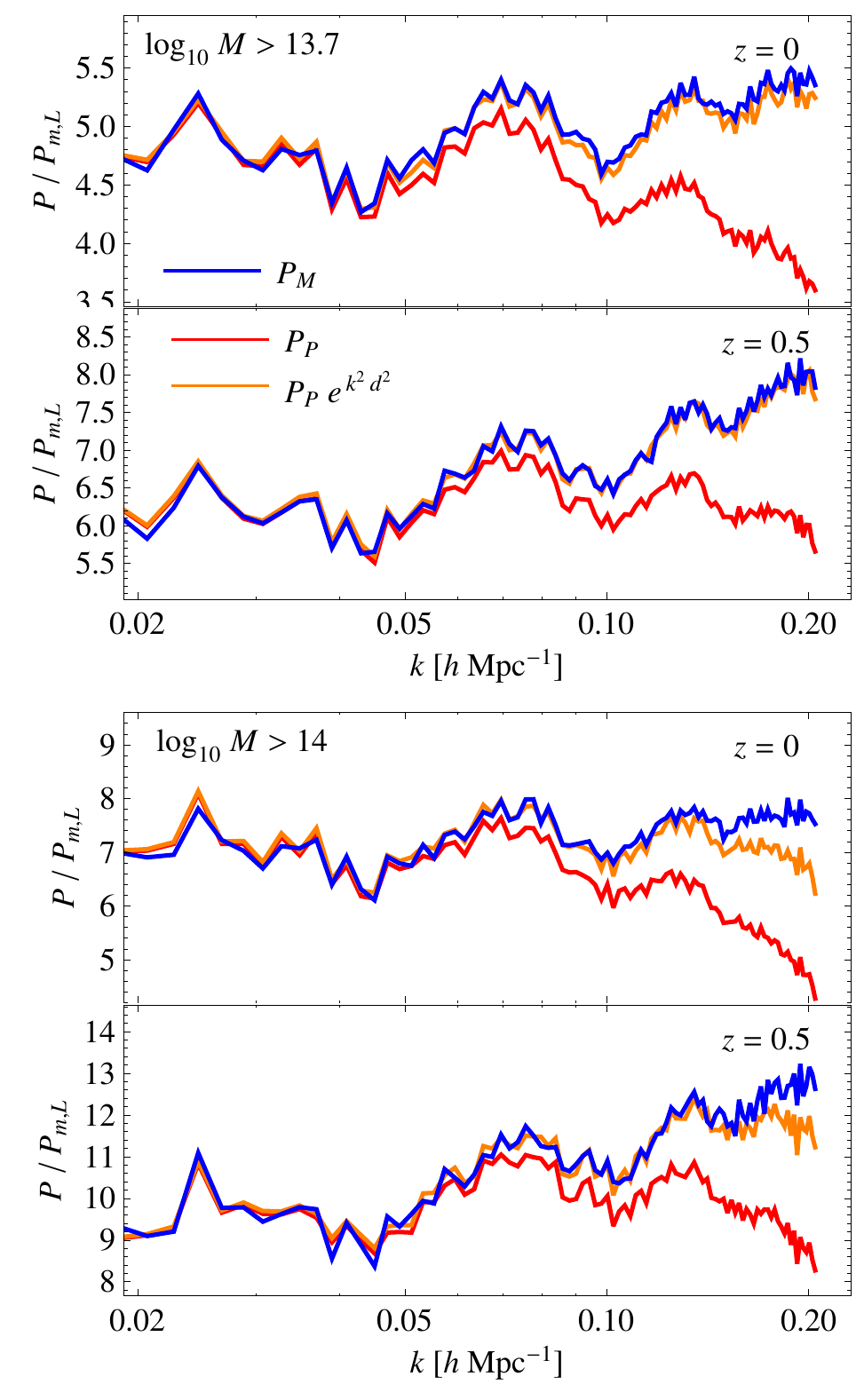}
\caption{\label{fig:bao} Comparison of the {\pin} and {\mice} halo power spectra over the acoustic oscillations range. We show the ratio of the {\miceB} ({\em blue curves}) and {\pinB} ({\em red curves}) power spectra w.r.t. the $\Omega_b=0$, linear matter power spectrum of \citet{EisensteinHu1998}. In addition, the orange curves show the {\pinB} power spectrum corrected according to Eq.~(\ref{expo}) with $d=3$ and $2.7\Mpc$ respectively at $z=0$ and $0.5$. The normalization of the {\pin} power spectrum is rescaled to match the {\mice} one at large scales to highlight the different scale dependence at small scales. Different panels show different thresholds in mass as indicated. In all cases, shot noise has been removed. {\pin} reproduces the sampling noise induced by the random initial conditions correctly over the whole range shown. }
\end{center}
\end{figure}

The inaccuracy of Zel'dovich displacements can be approximately described as a Gaussian scatter of halo positions about the true ones. Considering this uncertainty, the power spectrum of {\pin} halos, in turn, can be crudely modeled as the ``true'' one obtained from the simulation times a Gaussian, random scatter term:
\be 
P_{\rm\sc PIN}(k) = P_{\rm MICE}(k)\, e^{-k^2d^2}\, , \label{expo} 
\ee 
Fig.~\ref{fig:bao} shows the ratio of the {\mice} ({\em blue curves}) and {\pin} ({\em red curves}) power spectra (with shot-noise subtracted) w.r.t. the linear, matter power spectrum without acoustic oscillations \citep{EisensteinHu1998}. The normalization of the {\pin} power spectrum is rescaled to match the {\mice} one at large scales in order to highlight the different scale dependence at small scales. In addition, the figure shows a {\em corrected} {\pin} power spectrum obtained as $P_{\rm PIN}\,e^{k^2d^2}$, with $d=3$ and $2.7\Mpc$ respectively at $z=0$ and $0.5$.  The two panels show two different populations defined by different thresholds in mass, as indicated. After this Gaussian damping correction is applied, the residual difference is roughly a constant bias, whose value depends on mass, as already shown in Fig.~\ref{fig:psratio}. 

Note that the {\pin} halo power spectrum reproduces quite accurately 
the sampling noise 
of the N-body power spectrum over the whole Baryonic Acoustic
Oscillations (BAOs) range.  This is particularly evident from the
middle panel of Fig.~\ref{fig:bao}, corresponding to $\log_{10}
M>13.7$, where the corrected {\pin} values practically coincide with
the {\mice} measurements, including the constant bias term.  However,
additional corrections are required at smaller scales ($k>0.17\kMpc$).

\subsection{Bispectrum}
\label{bispectrum}

To provide a complete statistical description of the halo distribution 
it is necessary to consider its non-Gaussian properties. In this respect, 
the lowest order correlation statistic that measures this 
in Fourier space is the bispectrum. Here we compare measurements 
from {\miceB} and {\pinB} of the {\em reduced} halo bispectrum, 
$Q_h(k_1,k_2,k_3)$ \citep{Fry1984}. This quantity is defined as the 
ratio between the halo bispectrum, $B_h(k_1,k_2,k_3)$,
i.e. the three point function of the halo density field in Fourier
space, and a suitable combination of quadratic terms of the halo 
power spectrum, that is
\be\label{eq:redbisp}
Q_h(k_1,k_2,k_3)=\frac{B_h(k_1,k_2,k_3)}{P_h(k_1)\,P_h(k_2)+2~{\rm
    perm.}}\,.  
\ee 
The denominator removes the overall scale dependence of the halo 
bispectrum to highlight its dependence on the shape of the triangular 
configuration considered.

We measure the halo bispectrum $B(k_1,k_2,k_3)$ for all triangular
configurations defined by wavenumbers $k_i$ in bins of 
$\Delta k=0.004\kMpc$ up to a maximum value of $k_{max}=0.13\kMpc$, 
focusing therefore on large scales. On such scales, it is possible 
to approximate the halo bispectrum by \citep{FryGaztanaga1993} 
%
\be\label{eq:redbispg}
Q_h(k_1,k_2,k_3)=\frac1{b_1}\,Q(k_1,k_2,k_3)+\frac{b_2}{b_1^2}\,, \ee
where $Q(k_1,k_2,k_3)$ denotes the reduced bispectrum of the matter
distribution, defined similarly to Eq.~\ref{eq:redbisp} in terms of
matter correlators while $b_1$ and $b_2$ represent respectively the
linear and quadratic bias coefficient, assumed to be constant for a
given halo population \citep[for a recent assessment of the validity 
of this approximation see, for instance,][]{PollackSmithPorciani2012, ChanScoccimarroSheth2012, SefusattiCrocceDesjacques2010, SefusattiCrocceDesjacques2012}.

\begin{figure}
\begin{center}
\includegraphics[width=.48\textwidth]{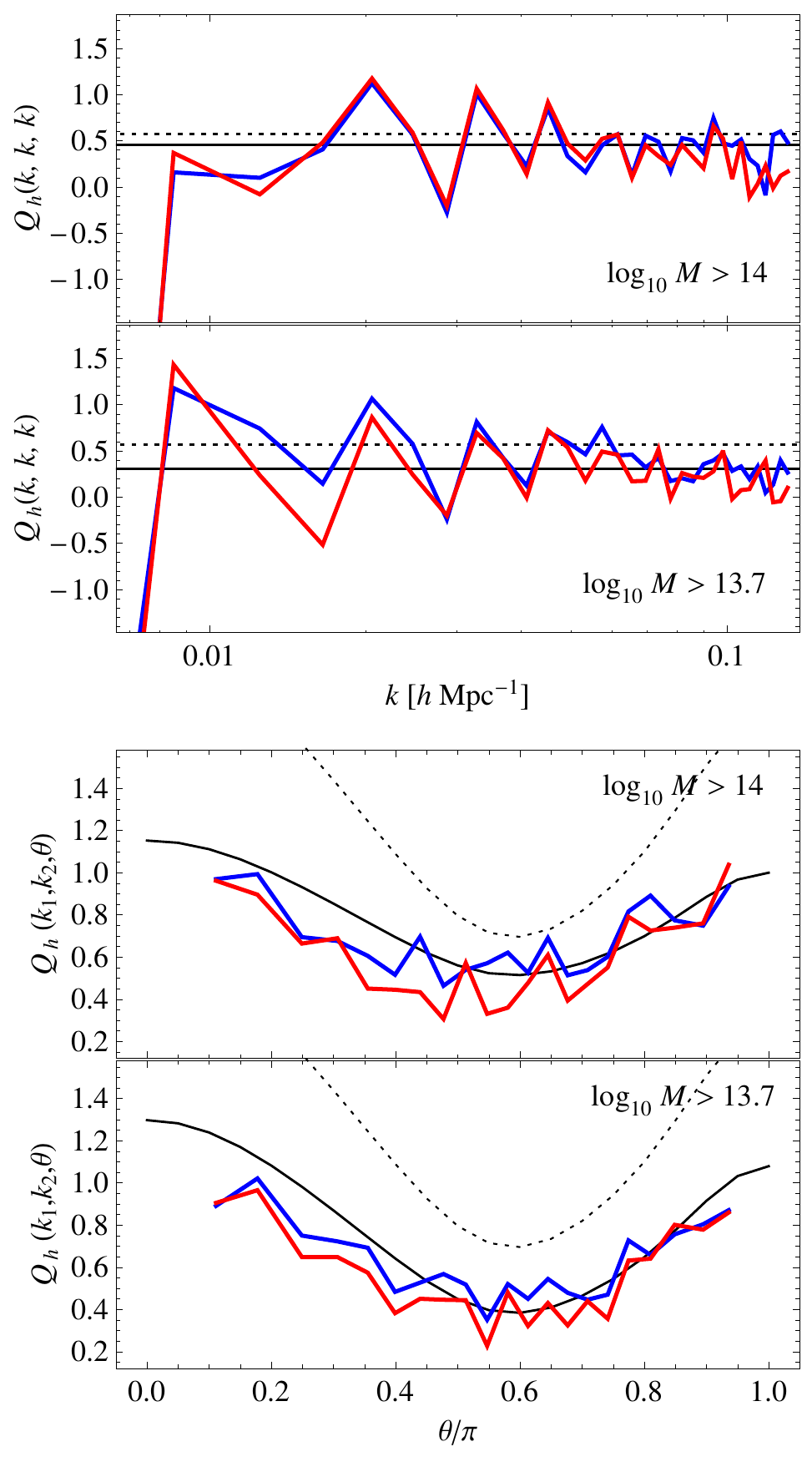}
\caption{\label{fig:bisp} 
Comparison of the reduced halo bispectrum. Top two panels show, for
two different mass thresholds, measurements of the reduced halo
bispectrum for equilateral configurations, $Q(k,k,k)$ as function of
$k$. Bottom two panel show the same quantity, $Q(k_1,k_2,\theta)$
with two wavenumbers are fixed to the values $k_1=0.04\kMpc$ and
$k_2=2 k_1$, as a function of the angle theta between them. In all
panels blue curves show measurements from {\miceB} and red curves
measurements from {\pinB}. The dotted, black curve provides the
prediction for the reduced {\em matter} bispectrum at tree-level in
perturbation theory. }
\end{center}
\end{figure}

From the expression above it is therefore evident that the reduced
halo bispectrum is equally sensitive to linear and nonlinear bias,
providing a test for the ability of {\pin} to correctly reproduce halo
bias beyond its linear approximation relevant for the large-scale
power spectrum. With this in mind, the upper panel of
Fig.~\ref{fig:bisp} shows the equilateral configurations of the
reduced halo bispectrum, $Q_h(k,k,k)$ as a function of $k$ for two
different mass thresholds at $z=0$. The bottom panel shows instead the
quantity $Q(k_1,k_2,\theta)$ with two wavenumbers fixed to the values
$k_1=0.04\kMpc$ and $k_2=2 k_1$, as a function of the angle $\theta$
between them. In all panels blue curves show measurements from
{\miceB} and red curves measurements from {\pinB}. The dotted, black
curve represents the prediction for the reduced {\em matter} bispectrum
at tree-level in perturbation theory, while the continuous black curve
shows Eq.~(\ref{eq:redbispg}), with the values for $b_1$ and $b_2$ 
determined from applying the peak-background split argument to the 
\citet{CrocceEtal2010} mass function: $b_1\simeq2.2$ and $2.6$ and 
$b_2\simeq 1$ and $2.2$ for the two mass thresholds of 
$\log_{10}M=13.7$ and 14 at $z=0$.
These theoretical predictions turn out to be rather accurate, despite 
the crudeness of the tree-level expression in Eq.~(\ref{eq:redbispg}).

These measurements show that {\pin} accurately reproduces the
hierarchical relation between halo power spectrum and bispectrum and
to a certain extent the nonlinear properties of halo bias. In fact,
although the bispectrum itself suffers from the same Gaussian damping 
effect which we saw for the halo-halo power spectrum, this approximately 
cancels out in the ratio defined in Eq.~(\ref{eq:redbisp}).  
This is remarkable also because the 1st-order, i.e. Zel'dovich, 
displacements do not guarantee good recovery of the nonlinear 
bispectrum, failing to reproduce even its tree-level expression in 
Eulerian PT.

\section{Discussion and conclusions}
\label{conclusions}

We have presented a new parallel (MPI) version of the {\pin} code, 
optimized to run on hundreds of cores of a super-computer.  
We showed that the {\pin} code can quickly produce simulated catalogs 
of DM halos that closely reproduce the results of very large
simulations like those of the {\mice} project \citep{CrocceEtal2010}:  
Halo abundances are almost universal, and within a few per cent of 
the fit proposed by \citet{WarrenEtal2006}.  On the other hand {\pin} 
underproduces the abundance of the rarest FoF objects found in some 
recent simulations, including those of the {\mice} group.  
This may indicate problems with the fragmentation part of the {\pin} 
code, but it may also be related to the tendency of the FOF algorithm 
to overlink halos.

In addition to the abundances, we showed that {\pin} can also
reproduce the spatial distribution of the halos.  Despite the fact
that it uses Zel'dovich displacements to compute the final positions
of halos, {\pin} can reproduce the matter-halo and the halo-halo power
spectra at $k<0.1\kMpc$.  The linear bias factor is well recovered as
long as the number density of objects is well matched\footnote{In
  a forthcoming paper \citep{ParanjapeEtal2013B} we will compare {\pin}
  predictions with simulations and theoretical expectations based on
  excursion set theory, and will show that {\pin} can reproduce the
  bias of DM halos well beyond the linear bias approximation.}.  Using mass
thresholds so as to match the same number density of halos in the
{\pin} and {\mice} catalogs, the matter-halo and the halo-halo power
spectra of the simulation are recovered to within 4 and 8 per cent
respectively.  At smaller scales the {\pin} power spectrum shows a
damping that is due to the inaccuracy of Zel'dovich displacements in
predicting the final positions of halos.  This can be roughly modeled
as a Gaussian noise term with a damping scale of $3\Mpc$ at $z=0$.
Good agreement is also obtained for the reduced bispectrum of the
halos, with the effects of the damping term approximately cancelling
out in the ratio.
For both 2-point and 3-point statistics the noise in the quantities 
computed from {\pin} catalogs closely follows that computed from 
simulations.
We did not address in this paper velocity fields.  This is an
  important step in the analysis that must be performed before extending
  the comparison to the redshift space. \citet{MonacoEtal2005} and
  \citet{HeisenbergSchaferBartelman2011} already presented studies of
  the behaviour of peculiar velocities of {\pin} halos.  A detailed
  study of velocity power spectrum and redshift-space clustering will
  be the subject of a forthcoming paper.

This version of the {\pin} code is particularly suited for addressing 
the massive production of catalogs of DM halos, the first step in the 
generation of mock galaxy catalogs using 
Halo Occupation Distribution, abundance matching or semi-analytic models.  
Its scaling properties demonstrate the feasibility of running many 
($\sim10000$) massive simulations in a reasonable amount of time:  
we estimate $2\times10^6$ CPU hours to produce $10000$ $2160^3$ 
boxes, though some minor parts of the code scale poorly and must be 
improved.  The typical result of a {\pin} run is a catalog of halos with 
known mass, position, velocity and merger history that requires orders 
of magnitude less disc space than needed by a typical simulation, 
not to mention the complicated and ill-defined post-processing needed
by a standard simulation to produce well-behaved halo merger histories, 
which are a natural outcome of {\pin}.  The speed-up comes at the cost 
of information about the internal structures of halos.  But with 
refined models of the evolution of DM halos after mergers, such as are 
commonly used in semi-analytic models to predict the merging time of
galaxies, it is possible to approximately reconstruct the abundance of 
halo substructures from halo merging histories
 \citep[see, e.g.,][]{GiocoliEtal2010}.

Clearly, {\pin} is not meant to be a substitute for N-body
simulations. Rather, a natural application of our code is the
determination of the covariance properties of large-scale structure
observables (e.g. the galaxy power spectrum), as well as the study of
systematic effects (e.g. the selection function) and possible
correlation between the two \citep[see, for instance,
][]{RossEtal2012}. Indeed, the mocks produced by
\citet{ManeraEtal2013} with a version of the {\pthalos} code for the
BOSS survey \citep{EisensteinEtal2011} were an essential ingredient
for many analyses beyond error estimation, like power spectrum at
large scales, BAOs, redshift distorsions.  {\pin} itself has been used
by \cite{DeLaTorreEtal2013} for computing the covariance matrix of the
redshift-space galaxy correlation function in the range of scales
$\sim1$ to $\sim30$ Mpc. This is a difficult range to reproduce,
because Zel'dovich displacements are inaccurate at these scales.
Nevertheless the authors could take advantage of a relatively large
number of {\pin} realizations by applying the shrinkage method of
\cite{PopeSzapudi2008}, using only a few mocks from the MultiDark
simulation by \cite{PradaEtal2012} to subtract out the bias in the
determination of the correlation function. While the determination of
uncertainties does not require percent accuracy, a very large number
of mock catalogs is crucial for the proper estimation of large
covariance matrices.

{\pin} shows several advantages compared to other 
simplified tools for the quick production of large-scale structure.  
Algorithms based on LPT to reproduce the non-linear matter density 
field may be quicker, but they are not as precise in determining where 
the DM halos are, especially at small masses \citep{ManeraEtal2013}.  
Methods which use Particle-Mesh integrations in a few time steps can 
be very accurate in the generating the nonlinear mass field 
\citep{TassevZaldarriagaEisenstein2013}, but the price paid is poor 
time sampling of halo merger histories, as well as the post-processing 
needed to produce halo catalogs in the first place.  The sparse 
time-sampling also complicates the generation of halo catalogs along 
the past light cone which is much simpler in {\pin} because all 
displacements are always done in one single time-step, so any level 
of time sampling can be easily achieved. 
In particular, masses are updated every time a particle is added to the 
group, and merger histories report masses for each merging pair of 
halos, so with the minimal output given by {\pin}, halo mass accretion
histories are available at each halo merger even without outputting
the halo catalogs many times.

The present version of the code works for a range of $\Lambda$CDM 
cosmologies that includes arbitrary redshift-dependent equation of 
state of the quintessence, and can be easily extended to non-Gaussian 
cosmologies simply by changing the initial conditions generator.  
We are currently developing {\pin} in two further directions.  
Positions of halos can be computed with 2nd and 3rd-order LPT with 
associated overheads in memory and CPU time amounting to $\sim30$ 
and $\sim100$ per cent.  We expect that 2LPT would improve the accuracy
with which halo positions (and hence masses) are predicted, thus 
improving the halo power spectra and bispectra (i.e., reducing the
corrections currently needed at large wavenumbers).  It also could 
help in recovering the right number density of very rare halos.  
Full 3LPT would allow one to predict the collapse times without using 
the ellipsoidal truncation of LPT proposed by \cite{Monaco1997A}, that
works under the approximation that using the growing mode as a time
coordinate factorizes the dependence of cosmology out of the dynamics
of a mass element.  This would allow one to quickly produce simulations 
in any cosmology where an LPT expansion can be formulated.

The other direction of development is the on-the-fly production of the
output on the past light cone of an observer randomly placed in the
simulation volume, taking advantage of the periodic boundary
conditions to replicate simulate a very large volume.  The fine time
sampling of {\pin} eliminates the need to output the full catalogs
many times as must be done if one wishes to reconstruct the past-light
cone at the post-processing level in the conventional way.

Our final aim is to propose a quick, flexible, scalable and
open-source tool to generate, with minimal resources, large catalogs of
DM halos that reproduce the statistics of simulations to an accuracy
which justifies the use of this tool for high-precision cosmology.

\section*{Acknowledgements}

We thank Stefano Anselmi for discussions.
P.M. and S.B. acknowledge financial contributions from the European
Commissions FP7 Marie Curie Initial Training Network CosmoComp
(PITN-GA-2009-238356), from PRIN MIUR 2010-2011 J91J12000450001 ``The
dark Universe and the cosmic evolution of baryons: from current
surveys to Euclid'', from PRIN-INAF 2009 ``Towards an Italian Network
for Computational Cosmology'', from ASI/INAF agreement I/023/12/0,
from PRIN-MIUR09 ``Tracing the growth of structures in the Universe''
and from a FRA2012 grant of the Trieste University. E.S. and R.S. were
supported in part by NSF-AST 0908241.

\bibliography{Bibliography}

\begin{thebibliography}{68}
\expandafter\ifx\csname natexlab\endcsname\relax\def\natexlab#1{#1}\fi

\bibitem[{{Ade} {et~al.}(2013{\natexlab{a}}){Ade}, {Aghanim},
  {Armitage-Caplan}, {Arnaud}, {Ashdown}, {Atrio-Barandela}, {Aumont},
  {Baccigalupi}, {Banday}, \& et~al.}]{Planck2013overview}
{Ade}, P.~A.~R. {et~al.} 2013{\natexlab{a}}, ArXiv e-prints, 1303.5062

\bibitem[{{Ade} {et~al.}(2013{\natexlab{b}}){Ade}, {Aghanim},
  {Armitage-Caplan}, {Arnaud}, {Ashdown}, {Atrio-Barandela}, {Aumont},
  {Baccigalupi}, {Banday}, \& et~al.}]{Planck2013parameters}
------. 2013{\natexlab{b}}, ArXiv e-prints, 1303.5076

\bibitem[{{Amendola} {et~al.}(2012){Amendola}, {Appleby}, {Bacon}, {Baker},
  {Baldi}, {Bartolo}, {Blanchard}, {Bonvin}, {Borgani}, {Branchini}, {Burrage},
  {Camera}, {Carbone}, {Casarini}, {Cropper}, {deRham}, {di Porto}, {Ealet},
  {Ferreira}, {Finelli}, {Garcia-Bellido}, {Giannantonio}, {Guzzo}, {Heavens},
  {Heisenberg}, {Heymans}, {Hoekstra}, {Hollenstein}, {Holmes}, {Horst},
  {Jahnke}, {Kitching}, {Koivisto}, {Kunz}, {La Vacca}, {March}, {Majerotto},
  {Markovic}, {Marsh}, {Marulli}, {Massey}, {Mellier}, {Mota}, {Nunes},
  {Percival}, {Pettorino}, {Porciani}, {Quercellini}, {Read}, {Rinaldi},
  {Sapone}, {Scaramella}, {Skordis}, {Simpson}, {Taylor}, {Thomas}, {Trotta},
  {Verde}, {Vernizzi}, {Vollmer}, {Wang}, {Weller}, \&
  {Zlosnik}}]{AmendolaEtal2012}
{Amendola}, L. {et~al.} 2012, ArXiv e-prints, 1206.1225

\bibitem[{Angulo {et~al.}(2012)Angulo, Springel, White, Jenkins, Baugh, \&
  Frenk}]{AnguloEtal2012}
Angulo, R.~E., Springel, V., White, S. D.~M., Jenkins, A., Baugh, C.~M., \&
  Frenk, C.~S. 2012, \mnras, 426, 2046, 1203.3216

\bibitem[{{Bennett} {et~al.}(2012){Bennett}, {Larson}, {Weiland}, {Jarosik},
  {Hinshaw}, {Odegard}, {Smith}, {Hill}, {Gold}, {Halpern}, {Komatsu}, {Nolta},
  {Page}, {Spergel}, {Wollack}, {Dunkley}, {Kogut}, {Limon}, {Meyer}, {Tucker},
  \& {Wright}}]{BennettEtal2012}
{Bennett}, C.~L. {et~al.} 2012, ArXiv e-prints, 1212.5225

\bibitem[{Benson {et~al.}(2012)Benson, Borgani, De~Lucia, Boylan-Kolchin, \&
  Monaco}]{BensonEtal2012}
Benson, A.~J., Borgani, S., De~Lucia, G., Boylan-Kolchin, M., \& Monaco, P.
  2012, \mnras, 419, 3590, 1107.4098

\bibitem[{{Bond} {et~al.}(1991){Bond}, {Cole}, {Efstathiou}, \&
  Kaiser}]{BondEtal1991}
{Bond}, J.~R., {Cole}, S., {Efstathiou}, G., \& Kaiser, N. 1991, \apj, 379, 440

\bibitem[{{Buchert} \& {Ehlers}(1993)}]{BuchertEhlers1993}
{Buchert}, T., \& {Ehlers}, J. 1993, \mnras, 264, 375

\bibitem[{Carbone {et~al.}(2012)Carbone, Fedeli, Moscardini, \&
  Cimatti}]{CarboneEtal2012}
Carbone, C., Fedeli, C., Moscardini, L., \& Cimatti, A. 2012, \jcap, 3, 23,
  1112.4810

\bibitem[{Catelan(1995)}]{Catelan1995}
Catelan, P. 1995, \mnras, 276, 115, arXiv:astro-ph/9406016

\bibitem[{Chan {et~al.}(2012)Chan, Scoccimarro, \&
  Sheth}]{ChanScoccimarroSheth2012}
Chan, K.~C., Scoccimarro, R., \& Sheth, R.~K. 2012, \prd, 85, 083509, 1201.3614

\bibitem[{{Costanzi Alunno Cerbolini} {et~al.}(2013){Costanzi Alunno
  Cerbolini}, Sartoris, Xia, {Biviano}, Borgani, \& Viel}]{CostanziEtal2013}
{Costanzi Alunno Cerbolini}, M., Sartoris, B., Xia, J.-Q., {Biviano}, A.,
  Borgani, S., \& Viel, M. 2013, ArXiv e-prints, 1303.4550

\bibitem[{{Courtin} {et~al.}(2011){Courtin}, {Rasera}, Alimi, Corasaniti,
  {Boucher}, \& {F{\"u}zfa}}]{CourtinEtal2011}
{Courtin}, J., {Rasera}, Y., Alimi, J.-M., Corasaniti, P., {Boucher}, V., \&
  {F{\"u}zfa}, A. 2011, \mnras, 410, 1911, 1001.3425

\bibitem[{Crocce {et~al.}(2010)Crocce, Fosalba, Castander, \&
  Gazta{\~n}aga}]{CrocceEtal2010}
Crocce, M., Fosalba, P., Castander, F.~J., \& Gazta{\~n}aga, E. 2010, \mnras,
  403, 1353, 0907.0019

\bibitem[{Crocce {et~al.}(2006)Crocce, Pueblas, \&
  Scoccimarro}]{CroccePueblasScoccimarro2006}
Crocce, M., Pueblas, S., \& Scoccimarro, R. 2006, \mnras, 373, 369, arXiv:
  astro-ph/0606505

\bibitem[{{Das} {et~al.}(2013){Das}, {Louis}, {Nolta}, {Addison},
  {Battistelli}, {Bond}, {Calabrese}, {Devlin}, {Dicker}, {Dunkley},
  {D{\"u}nner}, {Fowler}, {Gralla}, {Hajian}, {Halpern}, {Hasselfield},
  {Hilton}, {Hincks}, {Hlozek}, {Huffenberger}, {Hughes}, {Irwin}, {Kosowsky},
  {Lupton}, {Marriage}, {Marsden}, {Menanteau}, {Moodley}, {Niemack}, {Page},
  {Partridge}, {Reese}, {Schmitt}, {Sehgal}, {Sherwin}, {Sievers}, {Spergel},
  {Staggs}, {Swetz}, {Switzer}, {Thornton}, {Trac}, \& {Wollack}}]{DasEtal2013}
{Das}, S. {et~al.} 2013, ArXiv e-prints, 1301.1037

\bibitem[{de~la Torre {et~al.}(2013)de~la Torre, Guzzo, Peacock, Branchini,
  {Iovino}, {Granett}, {Abbas}, {Adami}, {Arnouts}, {Bel}, {Bolzonella},
  {Bottini}, {Cappi}, {Coupon}, {Cucciati}, {Davidzon}, {De Lucia}, {Fritz},
  {Franzetti}, {Fumana}, {Garilli}, {Ilbert}, {Krywult}, {Le Brun}, {Le Fevre},
  {Maccagni}, {Malek}, {Marulli}, {McCracken}, {Moscardini}, {Paioro},
  {Percival}, {Polletta}, {Pollo}, {Schlagenhaufer}, {Scodeggio}, {Tasca},
  {Tojeiro}, {Vergani}, {Zanichelli}, {Burden}, {Di Porto}, {Marchetti},
  {Marinoni}, {Mellier}, {Monaco}, {Nichol}, {Phleps}, {Wolk}, \&
  {Zamorani}}]{DeLaTorreEtal2013}
de~la Torre, S. {et~al.} 2013, ArXiv e-prints, 1303.2622

\bibitem[{Desjacques \& Seljak(2010)}]{DesjacquesSeljak2010C}
Desjacques, V., \& Seljak, U. 2010, Classical and Quantum Gravity, 27, 124011,
  1003.5020

\bibitem[{Eisenstein \& Hu(1998)}]{EisensteinHu1998}
Eisenstein, D.~J., \& Hu, W. 1998, \apj, 496, 605, arXiv:astro-ph/9709112

\bibitem[{{Eisenstein} {et~al.}(2011){Eisenstein}, {Weinberg}, {Agol},
  {Aihara}, {Allende Prieto}, {Anderson}, {Arns}, {Aubourg}, {Bailey},
  {Balbinot}, \& et~al.}]{EisensteinEtal2011}
{Eisenstein}, D.~J. {et~al.} 2011, \aj, 142, 72, 1101.1529

\bibitem[{Fosalba {et~al.}(2008)Fosalba, Gazta{\~n}aga, Castander, \&
  Manera}]{FosalbaEtal2008}
Fosalba, P., Gazta{\~n}aga, E., Castander, F.~J., \& Manera, M. 2008, \mnras,
  391, 435, 0711.1540

\bibitem[{{Frigo} \& {Johnson}(2012)}]{FrigoJohnson2012}
{Frigo}, M., \& {Johnson}, S.~G. 2012, {FFTW: Fastest Fourier Transform in the
  West}, 1201.015, astrophysics Source Code Library

\bibitem[{Fry(1984)}]{Fry1984}
Fry, J.~N. 1984, \apj, 279, 499

\bibitem[{Fry \& Gazta{\~n}aga(1993)}]{FryGaztanaga1993}
Fry, J.~N., \& Gazta{\~n}aga, E. 1993, \apj, 413, 447, astro-ph/9302009

\bibitem[{Giocoli {et~al.}(2010)Giocoli, Tormen, Sheth, \& van~den
  Bosch}]{GiocoliEtal2010}
Giocoli, C., Tormen, G., Sheth, R.~K., \& van~den Bosch, F.~C. 2010, \mnras,
  404, 502, 0911.0436

\bibitem[{{Heisenberg} {et~al.}(2011){Heisenberg}, {Sch{\"a}fer}, \&
  {Bartelmann}}]{HeisenbergSchaferBartelman2011}
{Heisenberg}, L., {Sch{\"a}fer}, B.~M., \& {Bartelmann}, M. 2011, \mnras, 416,
  3057, 1011.1559

\bibitem[{{Hinshaw} {et~al.}(2012){Hinshaw}, {Larson}, {Komatsu}, {Spergel},
  {Bennett}, {Dunkley}, {Nolta}, {Halpern}, {Hill}, {Odegard}, {Page}, {Smith},
  {Weiland}, {Gold}, {Jarosik}, {Kogut}, {Limon}, {Meyer}, {Tucker}, {Wollack},
  \& {Wright}}]{HinshawEtal2012}
{Hinshaw}, G. {et~al.} 2012, ArXiv e-prints, 1212.5226

\bibitem[{Jahnke \& Macci{\`o}(2011)}]{JahnkeMaccio2011}
Jahnke, K., \& Macci{\`o}, A.~V. 2011, \apj, 734, 92, 1006.0482

\bibitem[{{Jenkins} {et~al.}(2001){Jenkins}, {Frenk}, {White}, {Colberg},
  {Cole}, {Evrard}, {Couchman}, \& {Yoshida}}]{JenkinsEtal2001}
{Jenkins}, A., {Frenk}, C.~S., {White}, S.~D.~M., {Colberg}, J.~M., {Cole}, S.,
  {Evrard}, A.~E., {Couchman}, H.~M.~P., \& {Yoshida}, N. 2001, \mnras, 321,
  372, arXiv:astro-ph/0005260

\bibitem[{{Kitaura} \& {He{\ss}}(2012)}]{KitauraHess2012}
{Kitaura}, F.-S., \& {He{\ss}}, S. 2012, ArXiv e-prints, 1212.3514

\bibitem[{{Lahav} {et~al.}(2010){Lahav}, {Kiakotou}, {Abdalla}, \&
  {Blake}}]{LahavEtal2010}
{Lahav}, O., {Kiakotou}, A., {Abdalla}, F.~B., \& {Blake}, C. 2010, \mnras,
  405, 168, 0910.4714

\bibitem[{{Laureijs} {et~al.}(2011){Laureijs}, {Amiaux}, {Arduini},
  {Augu{\`e}res}, {Brinchmann}, {Cole}, {Cropper}, {Dabin}, {Duvet}, {Ealet},
  \& et~al.}]{LaureijsEtal2011}
{Laureijs}, R. {et~al.} 2011, ArXiv e-prints, 1110.3193

\bibitem[{{Li} {et~al.}(2007){Li}, {Mo}, {van den Bosch}, \&
  {Lin}}]{LiEtal2007}
{Li}, Y., {Mo}, H.~J., {van den Bosch}, F.~C., \& {Lin}, W.~P. 2007, \mnras,
  379, 689, arXiv:astro-ph/0510372

\bibitem[{Liguori {et~al.}(2010)Liguori, Sefusatti, Fergusson, \&
  Shellard}]{LiguoriEtal2010}
Liguori, M., Sefusatti, E., Fergusson, J.~R., \& Shellard, E. P.~S. 2010,
  Advances in Astronomy, 2010, 1001.4707

\bibitem[{{Lu} {et~al.}(2006){Lu}, {Mo}, {Katz}, \& {Weinberg}}]{LuEtal2006}
{Lu}, Y., {Mo}, H.~J., {Katz}, N., \& {Weinberg}, M.~D. 2006, \mnras, 368,
  1931, arXiv:astro-ph/0508624

\bibitem[{Manera {et~al.}(2013)Manera, Scoccimarro, Percival, {Samushia},
  McBride, Ross, Sheth, {White}, {Reid}, {S{\'a}nchez}, {de Putter}, {Xu},
  {Berlind}, {Brinkmann}, {Maraston}, {Nichol}, {Montesano}, {Padmanabhan},
  {Skibba}, {Tojeiro}, \& {Weaver}}]{ManeraEtal2013}
Manera, M. {et~al.} 2013, \mnras, 428, 1036, 1203.6609

\bibitem[{Merson {et~al.}(2013)Merson, Baugh, Helly, Gonzalez-Perez, Cole,
  Bielby, Norberg, Frenk, Benson, Bower, Lacey, \& Lagos}]{MersonEtal2013}
Merson, A.~I. {et~al.} 2013, \mnras, 429, 556, 1206.4049

\bibitem[{{Monaco}(1995)}]{Monaco1995}
{Monaco}, P. 1995, \apj, 447, 23, arXiv:astro-ph/9406029

\bibitem[{Monaco(1997)}]{Monaco1997A}
Monaco, P. 1997, \mnras, 287, 753, arXiv:astro-ph/9606027

\bibitem[{Monaco {et~al.}(2005)Monaco, {M{\o}ller}, {Fynbo}, {Weidinger},
  {Ledoux}, \& Theuns}]{MonacoEtal2005}
Monaco, P., {M{\o}ller}, P., {Fynbo}, J.~P.~U., {Weidinger}, M., {Ledoux}, C.,
  \& Theuns, T. 2005, \aap, 440, 799, arXiv:astro-ph/0505477

\bibitem[{Monaco {et~al.}(2002)Monaco, {Theuns}, \&
  {Taffoni}}]{MonacoTheunsTaffoni2002}
Monaco, P., {Theuns}, T., \& {Taffoni}, G. 2002, \mnras, 331, 587,
  arXiv:astro-ph/0109323

\bibitem[{{Moutarde} {et~al.}(1991){Moutarde}, {Alimi}, {Bouchet}, {Pellat}, \&
  {Ramani}}]{MoutardeEtal1991}
{Moutarde}, F., {Alimi}, J.-M., {Bouchet}, F.~R., {Pellat}, R., \& {Ramani}, A.
  1991, \apj, 382, 377

\bibitem[{Paranjape {et~al.}(2013)Paranjape, Sefusatti, Sheth, \&
  Monaco}]{ParanjapeEtal2013B}
Paranjape, A. {et~al.} 2013, in preparation

\bibitem[{Peel {et~al.}(2009)Peel, Battye, \& Kay}]{PeelBattyeKay2009}
Peel, M.~W., Battye, R.~A., \& Kay, S.~T. 2009, \mnras, 397, 2189, 0903.5473

\bibitem[{{Peel} {et~al.}(2009){Peel}, {Battye}, \& {Kay}}]{PeelEtal2009}
{Peel}, M.~W., {Battye}, R.~A., \& {Kay}, S.~T. 2009, \mnras, 397, 2189,
  0903.5473

\bibitem[{{Pierre} {et~al.}(2011){Pierre}, {Pacaud}, {Juin}, {Melin}, Valageas,
  {Clerc}, \& Corasaniti}]{PierreEtal2011}
{Pierre}, M., {Pacaud}, F., {Juin}, J.~B., {Melin}, J.~B., Valageas, P.,
  {Clerc}, N., \& Corasaniti, P. 2011, \mnras, 414, 1732, 1009.3182

\bibitem[{Pollack {et~al.}(2012)Pollack, Smith, \&
  Porciani}]{PollackSmithPorciani2012}
Pollack, J.~E., Smith, R.~E., \& Porciani, C. 2012, \mnras, 420, 3469,
  1109.3458

\bibitem[{Pope \& Szapudi(2008)}]{PopeSzapudi2008}
Pope, A.~C., \& Szapudi, I. 2008, \mnras, 389, 766, 0711.2509

\bibitem[{Prada {et~al.}(2012)Prada, Klypin, Cuesta, Betancort-Rijo, \&
  Primack}]{PradaEtal2012}
Prada, F., Klypin, A.~A., Cuesta, A.~J., Betancort-Rijo, J.~E., \& Primack,
  J.~R. 2012, \mnras, 423, 3018, 1104.5130

\bibitem[{Reed {et~al.}(2007)Reed, {Bower}, Frenk, Jenkins, \&
  Theuns}]{ReedEtal2007}
Reed, D.~S., {Bower}, R., Frenk, C.~S., Jenkins, A., \& Theuns, T. 2007,
  \mnras, 374, 2, arXiv:astro-ph/0607150

\bibitem[{{Ross} {et~al.}(2012){Ross}, {Percival}, {S{\'a}nchez}, {Samushia},
  {Ho}, {Kazin}, {Manera}, {Reid}, {White}, {Tojeiro}, {McBride}, {Xu}, {Wake},
  {Strauss}, {Montesano}, {Swanson}, {Bailey}, {Bolton}, {Dorta}, {Eisenstein},
  {Guo}, {Hamilton}, {Nichol}, {Padmanabhan}, {Prada}, {Schlegel},
  {Maga{\~n}a}, {Zehavi}, {Blanton}, {Bizyaev}, {Brewington}, {Cuesta},
  {Malanushenko}, {Malanushenko}, {Oravetz}, {Parejko}, {Pan}, {Schneider},
  {Shelden}, {Simmons}, {Snedden}, \& {Zhao}}]{RossEtal2012}
{Ross}, A.~J. {et~al.} 2012, \mnras, 424, 564, 1203.6499

\bibitem[{Schneider {et~al.}(2006)Schneider, Salvaterra, Ferrara, \&
  Ciardi}]{SchneiderEtal2006}
Schneider, R., Salvaterra, R., Ferrara, A., \& Ciardi, B. 2006, \mnras, 369,
  825, arXiv:astro-ph/0510685

\bibitem[{Scoccimarro {et~al.}(2012)Scoccimarro, Hui, Manera, \&
  Chan}]{ScoccimarroEtal2012}
Scoccimarro, R., Hui, L., Manera, M., \& Chan, K.~C. 2012, \prd, 85, 083002,
  1108.5512

\bibitem[{Sefusatti {et~al.}(2010)Sefusatti, Crocce, \&
  Desjacques}]{SefusattiCrocceDesjacques2010}
Sefusatti, E., Crocce, M., \& Desjacques, V. 2010, \mnras, 721, 1003.0007

\bibitem[{Sefusatti {et~al.}(2012)Sefusatti, Crocce, \&
  Desjacques}]{SefusattiCrocceDesjacques2012}
------. 2012, \mnras, 425, 2903, 1111.6966

\bibitem[{Sefusatti {et~al.}(2006)Sefusatti, Crocce, Pueblas, \&
  Scoccimarro}]{SefusattiEtal2006}
Sefusatti, E., Crocce, M., Pueblas, S., \& Scoccimarro, R. 2006, \prd, 74,
  023522, arXiv: astro-ph/0604505

\bibitem[{Sheth \& Tormen(1999)}]{ShethTormen1999}
Sheth, R.~K., \& Tormen, G. 1999, \mnras, 308, 119, astro-ph/9901122

\bibitem[{{Sievers} {et~al.}(2013){Sievers}, {Hlozek}, {Nolta}, {Acquaviva},
  {Addison}, {Ade}, {Aguirre}, {Amiri}, {Appel}, {Barrientos}, {Battistelli},
  {Battaglia}, {Bond}, {Brown}, {Burger}, {Calabrese}, {Chervenak}, {Crichton},
  {Das}, {Devlin}, {Dicker}, {Bertrand Doriese}, {Dunkley}, {D{\"u}nner},
  {Essinger-Hileman}, {Faber}, {Fisher}, {Fowler}, {Gallardo}, {Gordon},
  {Gralla}, {Hajian}, {Halpern}, {Hasselfield}, {Hern{\'a}ndez-Monteagudo},
  {Hill}, {Hilton}, {Hilton}, {Hincks}, {Holtz}, {Huffenberger}, {Hughes},
  {Hughes}, {Infante}, {Irwin}, {Jacobson}, {Johnstone}, {Baptiste Juin},
  {Kaul}, {Klein}, {Kosowsky}, {Lau}, {Limon}, {Lin}, {Louis}, {Lupton},
  {Marriage}, {Marsden}, {Martocci}, {Mauskopf}, {McLaren}, {Menanteau},
  {Moodley}, {Moseley}, {Netterfield}, {Niemack}, {Page}, {Page}, {Parker},
  {Partridge}, {Plimpton}, {Quintana}, {Reese}, {Reid}, {Rojas}, {Sehgal},
  {Sherwin}, {Schmitt}, {Spergel}, {Staggs}, {Stryzak}, {Swetz}, {Switzer},
  {Thornton}, {Trac}, {Tucker}, {Uehara}, {Visnjic}, {Warne}, {Wilson},
  {Wollack}, {Zhao}, \& {Zuncke}}]{SieversEtal2013}
{Sievers}, J.~L. {et~al.} 2013, ArXiv e-prints, 1301.0824

\bibitem[{{Springel}(2005)}]{Springel2005}
{Springel}, V. 2005, \mnras, 364, 1105, arXiv:astro-ph/0505010

\bibitem[{{Springel} {et~al.}(2005){Springel}, {White}, {Jenkins}, {Frenk},
  {Yoshida}, {Gao}, {Navarro}, {Thacker}, {Croton}, {Helly}, {Peacock}, {Cole},
  {Thomas}, {Couchman}, {Evrard}, {Colberg}, \& {Pearce}}]{SpringelEtal2005}
{Springel}, V. {et~al.} 2005, \nat, 435, 629, arXiv:astro-ph/0504097

\bibitem[{{Story} {et~al.}(2012){Story}, {Reichardt}, {Hou}, {Keisler}, {Aird},
  {Benson}, {Bleem}, {Carlstrom}, {Chang}, {Cho}, {Crawford}, {Crites}, {de
  Haan}, {Dobbs}, {Dudley}, {Follin}, {George}, {Halverson}, {Holder},
  {Holzapfel}, {Hoover}, {Hrubes}, {Joy}, {Knox}, {Lee}, {Leitch}, {Lueker},
  {Luong-Van}, {McMahon}, {Mehl}, {Meyer}, {Millea}, {Mohr}, {Montroy},
  {Padin}, {Plagge}, {Pryke}, {Ruhl}, {Sayre}, {Schaffer}, {Shaw}, {Shirokoff},
  {Spieler}, {Staniszewski}, {Stark}, {van Engelen}, {Vanderlinde}, {Vieira},
  {Williamson}, \& {Zahn}}]{StoryEtal2012}
{Story}, K.~T. {et~al.} 2012, ArXiv e-prints, 1210.7231

\bibitem[{{Taffoni} {et~al.}(2002){Taffoni}, Monaco, \&
  {Theuns}}]{TaffoniMonacoTheuns2002}
{Taffoni}, G., Monaco, P., \& {Theuns}, T. 2002, \mnras, 333, 623,
  arXiv:astro-ph/0109324

\bibitem[{{Tassev} {et~al.}(2013){Tassev}, {Zaldarriaga}, \&
  {Eisenstein}}]{TassevZaldarriagaEisenstein2013}
{Tassev}, S., {Zaldarriaga}, M., \& {Eisenstein}, D. 2013, ArXiv e-prints,
  1301.0322

\bibitem[{Tinker {et~al.}(2008)Tinker, Kravtsov, Klypin, Abazajian, {Warren},
  {Yepes}, {Gottl{\"o}ber}, \& {Holz}}]{TinkerEtal2008}
Tinker, J., Kravtsov, A.~V., Klypin, A.~A., Abazajian, K., {Warren}, M.~S.,
  {Yepes}, G., {Gottl{\"o}ber}, S., \& {Holz}, D.~E. 2008, \apj, 688, 709,
  0803.2706

\bibitem[{{Warren} {et~al.}(2006){Warren}, Abazajian, {Holz}, \&
  {Teodoro}}]{WarrenEtal2006}
{Warren}, M.~S., Abazajian, K., {Holz}, D.~E., \& {Teodoro}, L. 2006, \apj,
  646, 881, arXiv:astro-ph/0506395

\bibitem[{{Zhao} {et~al.}(2003){Zhao}, {Jing}, {Mo}, \&
  {B{\"o}rner}}]{ZhaoEtal2003}
{Zhao}, D.~H., {Jing}, Y.~P., {Mo}, H.~J., \& {B{\"o}rner}, G. 2003, \apjl,
  597, L9, arXiv:astro-ph/0309375

\bibitem[{{Zhao} {et~al.}(2009){Zhao}, {Jing}, {Mo}, \&
  {B{\"o}rner}}]{ZhaoEtal2009}
------. 2009, \apj, 707, 354, 0811.0828

\bibitem[{Zheng {et~al.}(2007)Zheng, Coil, \& Zehavi}]{ZhengCoilZehavi2007}
Zheng, Z., Coil, A.~L., \& Zehavi, I. 2007, \apj, 667, 760,
  arXiv:astro-ph/0703457

\end{thebibliography}

\label{lastpage}

\end{document}